\documentstyle[aps]{revtex}
\begin{document}
\input psfig.sty
\draft
\twocolumn[\hsize\textwidth\columnwidth\hsize\csname@twocolumnfalse%
\endcsname

\begin{flushright}
       {\bf UK/99-20}  \\
 \end{flushright}

\title{\bf Quark Orbital Angular Momentum from Lattice QCD}

\author{N. Mathur$^{a,b}$, S. J. Dong$^{a}$, K. F. Liu$^{a,c}$,
L. Mankiewicz$^{d}$, and N. C. Mukhopadhyay$^{b}$}

\address
{$^{a}$Dept. of Physics and Astronomy, University of Kentucky,
Lexington, KY 40506\\
$^{b}$Dept. of Physics, Applied Physics and Astronomy, 
RPI, Troy, NY 12180 \\
$^{c}$SLAC, P. O. Box 4349, Stanford, CA 94309 \\
$^{d}$Physics Dept., Technical Univ., Munich, D-85747, Garching, Germany}


\maketitle
\vskip 12pt

\begin{abstract}
We calculate the quark orbital angular momentum of the nucleon from
the quark energy-momentum tensor form factors on the lattice.
The disconnected insertion is estimated 
stochastically which employs the $Z_2$ noise with an unbiased subtraction. 
This reduced the error by a factor of 4 with negligible overhead. The total
quark contribution to the proton spin is found to be $0.30 \pm 0.07$. From
this and the quark spin content we deduce the quark orbital angular 
momentum to be $0.17 \pm 0.06$ which is $\sim 34$\% of the proton spin. 
We further predict that the gluon angular momentum to be $0.20 \pm 0.07$, 
i. e. $\sim$ 40\% of the proton spin is due to the glue.
\end{abstract}
]

  The spin content of the proton remains a challenging problem in
QCD both experimentally and theoretically~\cite{review}. The surprisingly small
contribution from the quark spin revealed by the polarized deep inelastic
scattering experiments~\cite{EMC88} (world average: $\Sigma = 0.25
\pm 0.10$) has stimulated a great deal of interest in the understanding of
this `proton spin problem'. While the lattice QCD calculations~\cite{dll95}
confirmed the small quark spin content in agreement with experiments,
there is little consensus on where the rest of the proton spin resides. There
have been suggestions based on the Bjorken sum rule~\cite{seh74}, the
parton evolution~\cite{rat87}, the
chiral quark model~\cite{cl98} and skyrmion~\cite{li94} that
the quark orbital angular momentum in the nucleon can be substantial. 
It is further proposed that the off-forward parton distributions from the
deeply virtual Compton scattering can be used to measure the quark orbital
angular momentum distribution and thereby its moments~\cite{ji97}. 

   In this letter, we shall report the first lattice calculation of the
quark energy-momentum tensor form factors which admits the extraction of 
the total quark angular momentum. Combining the lattice calculation of the 
quark spin content~\cite{dll95}, we obtain the quark orbital angular momentum 
and thereby predict the gluon angular momentum in the nucleon from the 
spin sum rule.
It turns out that the quark orbital angular momentum is indeed quite large.
It constitutes $\sim$ 35 \% of the proton spin and the gluon angular
momentum is
predicted to make up the remaining $\sim$ 40\% of the proton spin.


It has been shown recently~\cite{ji97} that the total angular momentum in 
QCD can be decomposed into three pieces in a {\it gauge invariant} way
\begin{eqnarray}  \label{ssr}
\vec{J}& = & \int d^{3}x \, {1 \over 2}\, \bar{\psi}\,
\vec{\gamma} \gamma_{5} \psi \,+\,\int d^{3}x \, 
\bar{\psi}\gamma_4 \{\vec{x} \times (- i \vec{D})\} \psi  \nonumber\\
&&\,+\, {\int d^{3}x \, [\,\vec{x} \times {(\vec{E} \times 
\vec{B})}]}\, =\, {1\over 2} \vec{\Sigma} + \vec{L}_{q} + \vec{J_{g}},
\end{eqnarray}
The quark spin contribution ${1\over 2} \Sigma$ 
is defined through the forward matrix element of
the flavor-singlet axial-vector current. Similarly, $L_{q}$, the quark 
orbital angular momentum, is defined from the operator $\vec{L}_{q}$ and
$J_{g}$, the gluon angular momentum, is defined from
$\vec{J_{g}}$ which is   
assocaited with the gluon Poynting vector $\vec{E} \times \vec{B}$. 
The total quark angular 
momentum is $J_{q} = {1\over 2} \Sigma + L_{q}$. Whereas, the total gluon 
angular momentum $J_{g}$ cannot be further decomposed into gluon spin and 
gluon orbital angular momentum without explicit gauge dependence.

$J_{q}$ can be obtained from the energy-momentum tensor 
form factors~\cite{ji97}. One can write the gauge invariant quark-gluon 
energy-momentum tensor as
\begin{eqnarray}
T_{\mu\nu} &= & T_{\mu\nu}^{q} + T_{\mu\nu}^{g} \nonumber\\
&= & {1 \over 2}\,  \bar{\psi}\gamma_{(\mu}
[{\stackrel {\longrightarrow} {D}} -
{\stackrel {\longleftarrow} {D}}]_{\nu)}\psi 
\,+\, F_{\mu\alpha}F_{\nu\alpha}\,-\,{1\over 4} \delta_{\mu\nu} F^{2},
\end{eqnarray}
where the first part is the quark energy-momentum tensor and the second one 
is that of the gluon. Form factors of the 
energy-momentum tensor for either the quark or the gluon are 
defined by~\cite{ji97},
\begin{eqnarray}   \label{ff}
&&<p,s|T_{\mu\nu}(0)|p^{\prime},s^{\prime}> =
\bar {u}(p,s)[ T_{1}(q^{2})\,i \gamma_{(\mu}\bar{p}_{\nu)}
\nonumber\\
&&\,\,\,\,-\, T_{2}(q^{2}) \bar{p}_{(\mu} i\sigma_{\nu ) 
\alpha}q_{\alpha} /2m_N \,+\, {1\over {m_N}} T_{3}(q^{2})
(q_{\mu}q_{\nu} -  \delta_{\mu\nu}q^{2}) \nonumber\\
&&\,\,\,\,-\,m_N\, T_{4}(q^{2}) \delta_{\mu\nu} ] 
u(p^{\prime},s^{\prime}),
\end{eqnarray}
where $\bar{p}_{\mu} = (p_{\mu} + {p^{\prime}}_{\mu})/2,\, 
q_{\mu} = p_{\mu} - {p^{\prime}}_{\mu}$ and $u(p)$ is the nucleon spinor.
It can be proven that, for polarized target, the total angular momentum of 
the quarks or gluons  
\begin{eqnarray}
J_{q,g}&=& 
<p,s|\,{1 \over 2}\, \epsilon^{ijk}
\int d^{3}x \,(T_{4k}^{q,g}x_{j} - T_{4j}^{q,g}x_{k})\,|p,s>
\over <p,s|p,s>\nonumber\\
&=& {1 \over 2}\, [ T^{q,g}_{1}(0)\,+\,T^{q,g}_{2}(0)].
\end{eqnarray}
Therefore, the total angular momentum of the
quarks/gluons can be calculated at the $q^{2} \rightarrow 0$ limit of
the form factor $1/2[T_1^{q,g}(q^{2}) +T_2^{q,g}(q^{2})] $.

In Eq. (\ref{ff}), the energy momentum tensor is expressed in terms of 
four form factors; whereas, for calculating the angular momentum we need only 
two, namely $T_{1}$ and $T_{2}$. $T_{4j}$, with $j$ in the 
3-direction, does not admit the $T_{4}$ term. To remove $T_{3}$, we choose 
the momentum transfer to be orthogonal to the $j$ direction. 
 In order to get the required $T_{1}(q^{2}) +
 T_{2}(q^{2})$, we calculate the three point function 
$G_{NT_{4 j}N}(t_{2},t_{1},\vec{p},-\vec{q})$ for the operator $T_{4j}$.
The three point function has two parts: connected insertion (CI), due to 
the valence and cloud quarks, and disconnected insertion (DI) attributed to 
the sea quarks~\cite{liu98} (Fig.1). 
\vspace{-0.36in}
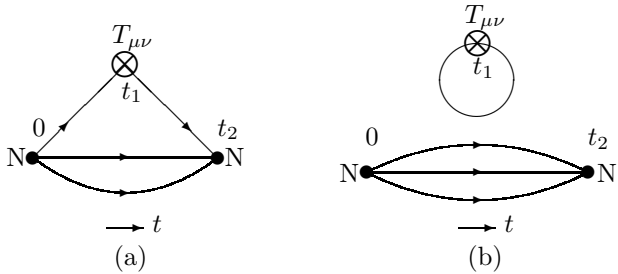
\begin{figure}[h]
\[
\hspace*{1.2in}
\setlength{\unitlength}{0.007pt}
\begin{picture}(45000,20000)
\put(-10000, 8800){\circle*{700}}
\put(-10000, 8800){\line(1,1){5000}}
\put(-8232, 10568){\vector(1,1){200}}
\put(0000, 8800){\line(-1,1){5000}}
\put(-1591,10391){\vector(1,-1){200}}
\put(-5800, 13500){{\bf $\bigotimes$}}
\put(-10000, 8800){\line(1,0){10000}}
\put(-5000,  8800){\vector(1,0){300}}
\put(0000, 8800){\circle*{700}}
\qbezier(-10000, 8800)(-5000, 5000)(00, 8800)
\put(-5000, 6900){\vector(1,0){300}}
\put(-6000, 5000){\vector(1,0){2000}}
\put(-5700,15000){{\bf $T_{\mu\nu}$}}
\put(-5200,12000){$t_1$}
\put(-3500, 4800){$t$}
\put(-10000,10000){$0$}
\put(0,10000){$t_{2}$}
\put(-11400,8350){N}
\put(400,8350){N}
\put(-5600, 3000){(a)}
\put(8050, 8000){\circle*{700}}
\put(8000, 8000){\line(1,0){12000}}
\put(14000, 8000){\vector(1,0){400}}
\put(20000, 8000){\circle*{700}}
\qbezier(8000, 8000)(14000, 5000)(20000, 8000)
\put(14000, 6500){\vector(1,0){400}}
\qbezier(8000, 8000)(14000,11000)(20000, 8000)
\put(14000, 9500){\vector(1,0){400}}
\put(14000,13000){\circle{4000}}
\put(13250,14600){{\bf $\bigotimes$}}
\put(13350,16100){{\bf $T_{\mu\nu}$}}
\put(13000, 5000){\vector(1,0){2000}}
\put(13800,13300){$t_1$}
\put(15500, 4800){$t$}
\put(8000, 9500){$0$}
\put(20000, 9500){$t_{2}$}
\put(6580,7500){N}
\put(20500,7500){N}
\put(13500, 3000){(b)}
%
\end{picture}
\]
\vspace{-0.4in}
\caption{Quark skeleton diagrams for the three-point function
$G_{NT_{4 j}N}(t_{2},t_{1},\vec{p},-\vec{q})$.
(a) is the connected insertion. (b) is the disconnected insertion.}
\end{figure}
For CI, we express the observable $T(q^{2}) \equiv {1 \over 2}\,
[T_1(q^2) + T_2(q^2)]$ in terms of 
the ratio of three- and two-point functions :
\begin{eqnarray} \label{ratio}
&&\hspace{-0.3in}{\hbox{Tr}\left[\Gamma_{m}G_{NT_{4 j}N}(t_{2},t_{1},\vec{0},
-\vec{q})\right]
\over {\hbox{Tr}}\left[\Gamma_{e}G_{NN}(t_{2}, \vec{0})\right]}\cdot 
{{\hbox{Tr}}\left[\Gamma_{e}G_{NN}(t_{1}, \vec{0})\right] \over 
{\hbox{Tr}}\left[\Gamma_{e}G_{NN}(t_{1}, \vec{q})\right]}\nonumber\\
&&=  {1\over 2}\,\epsilon_{jkm}\, q_{k}\, T(q^{2})\,,
\end{eqnarray}
where $\Gamma_{m}$ and $\Gamma_{e}$ are the spin polarized and
unpolarized projection operators~\cite{ldd94}.
From the above ratio, we calculate the lattice $T_{CI}(q^{2})$ at different 
$q^{2}$ and then extrapolate them to $q^{2} \rightarrow 0$ limit to 
obtain the CI part of the lattice quark total angular momentum $J_{q, CI}$. 
Results are obtained for relatively light Wilson quarks with
$\kappa\,=\,0.148, 0.152, 0.154$ and $0.155$ which correspond to 
quark masses of about $376, 210$, $124$ and $80$ MeV respectively.
Fig.2 shows the dipole fitting of $T_{CI}(q^{2})$ at different $q^{2}$.
Following the calculation for the 
point-split Wilson current~\cite{ldd94}, the tadpole improvement factor 
$1/8\kappa_{c}\langle {1\over3} {\hbox{Tr}}\, U_{plaq}\rangle
^{1/4}\, (\kappa_{c} = 0.15684)$ is included in the unrenormalized 
$J_{q, CI}$. The chiral limit for $J_{q, CI}$ is 
taken from a linear dependence on the quark mass $m_{q}a$ for 
these four $\kappa$ values (see Fig.3). To account for the correlations, 
both the dipole fitting and the chiral extrapolation 
are done with the covariance matrix and the final error at the chiral limit 
is obtained from the jackknife procedure. The dipole mass is found
to be $0.88 \pm 0.07\,GeV$. Finally, to obtain the 
result in the $\overline{MS}$ scheme at $1/a = 1.74\,GeV$, we multiply the  
$J_{q, CI}$ by the tadpole improved renormalization constant for the 
operator $T_{4j}$ which has been calculated~\cite{cap97} perturbatively 
to be $Z= 1.045$ and obtain the CI part of the quark angular momentum 
$J_{q, CI} = 0.44 \pm 0.07$. Being $\sim 90$\% of the nucleon spin, this 
almost saturates the spin sum rule. We also performed monopole fitting for the
$T_{CI}(q^2)$ and found an order of magnitude larger $\chi^{2}$.  
The calculation is done on a quenched $16^3 \times 24$ lattice 
at $\beta = 6.0$ with Wilson fermions for 100 configurations. 
\vspace*{-0.2in}
\begin{figure}[h]
\[
\hspace*{0.08in}
\input{spin_pole_c48m.tex}
\hspace*{-0.80in}
\input{spin_pole_c52m.tex}
\]
\end{figure}
\vspace{-1.9859cm}
\begin{figure}[h]
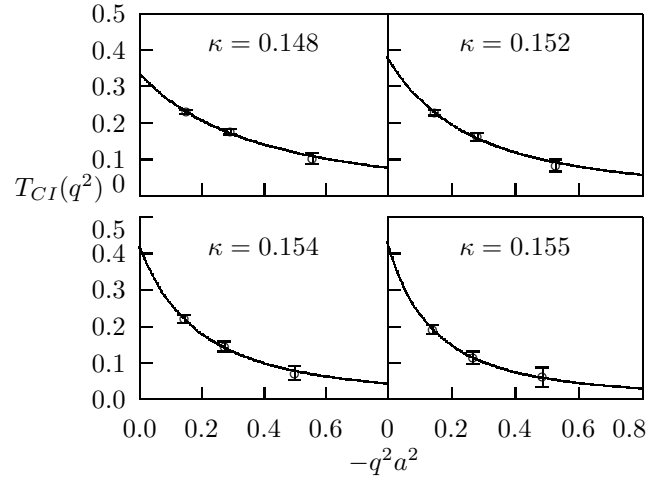

\[
\hspace*{0.08in}
\input{spin_pole_c54m.tex}
\hspace*{-0.80in}
\input{spin_pole_c55m.tex}
\]
\vspace{-0.1in}
\caption{Dipole fitting for $T_{CI}(q^2)$ at
4 different $\kappa$ values.}
\end{figure}
\vspace{-0.15in}
\begin{figure}[h]
\[
\hspace*{-0.3in}
\setlength{\unitlength}{0.240900pt}
\ifx\plotpoint\undefined\newsavebox{\plotpoint}\fi
\sbox{\plotpoint}{\rule[-0.200pt]{0.400pt}{0.400pt}}%
\begin{picture}(854,431)(0,0)
\font\gnuplot=cmr10 at 10pt
\gnuplot
\sbox{\plotpoint}{\rule[-0.200pt]{0.400pt}{0.400pt}}%
\put(220.0,113.0){\rule[-0.200pt]{4.818pt}{0.400pt}}
\put(198,113){\makebox(0,0)[r]{0}}
\put(770.0,113.0){\rule[-0.200pt]{4.818pt}{0.400pt}}
\put(220.0,187.0){\rule[-0.200pt]{4.818pt}{0.400pt}}
\put(198,187){\makebox(0,0)[r]{0.2}}
\put(770.0,187.0){\rule[-0.200pt]{4.818pt}{0.400pt}}
\put(220.0,261.0){\rule[-0.200pt]{4.818pt}{0.400pt}}
\put(198,261){\makebox(0,0)[r]{0.4}}
\put(770.0,261.0){\rule[-0.200pt]{4.818pt}{0.400pt}}
\put(220.0,334.0){\rule[-0.200pt]{4.818pt}{0.400pt}}
\put(198,334){\makebox(0,0)[r]{0.6}}
\put(770.0,334.0){\rule[-0.200pt]{4.818pt}{0.400pt}}
\put(220.0,408.0){\rule[-0.200pt]{4.818pt}{0.400pt}}
\put(198,408){\makebox(0,0)[r]{0.8}}
\put(770.0,408.0){\rule[-0.200pt]{4.818pt}{0.400pt}}
\put(220.0,113.0){\rule[-0.200pt]{0.400pt}{4.818pt}}
\put(220,68){\makebox(0,0){0}}
\put(220.0,388.0){\rule[-0.200pt]{0.400pt}{4.818pt}}
\put(315.0,113.0){\rule[-0.200pt]{0.400pt}{4.818pt}}
\put(315.0,388.0){\rule[-0.200pt]{0.400pt}{4.818pt}}
\put(410.0,113.0){\rule[-0.200pt]{0.400pt}{4.818pt}}
\put(410,68){\makebox(0,0){0.1}}
\put(410.0,388.0){\rule[-0.200pt]{0.400pt}{4.818pt}}
\put(505.0,113.0){\rule[-0.200pt]{0.400pt}{4.818pt}}
\put(505.0,388.0){\rule[-0.200pt]{0.400pt}{4.818pt}}
\put(600.0,113.0){\rule[-0.200pt]{0.400pt}{4.818pt}}
\put(600,68){\makebox(0,0){0.2}}
\put(600.0,388.0){\rule[-0.200pt]{0.400pt}{4.818pt}}
\put(695.0,113.0){\rule[-0.200pt]{0.400pt}{4.818pt}}
\put(695.0,388.0){\rule[-0.200pt]{0.400pt}{4.818pt}}
\put(790.0,113.0){\rule[-0.200pt]{0.400pt}{4.818pt}}
\put(778,68){\makebox(0,0){0.3}}
\put(790.0,388.0){\rule[-0.200pt]{0.400pt}{4.818pt}}
\put(220.0,113.0){\rule[-0.200pt]{137.313pt}{0.400pt}}
\put(790.0,113.0){\rule[-0.200pt]{0.400pt}{71.065pt}}
\put(220.0,408.0){\rule[-0.200pt]{137.313pt}{0.400pt}}
\put(65,260){\makebox(0,0){{$J_{q, CI}$}}}
\put(505,23){\makebox(0,0){{ $m_{q}a = ln(4\kappa_{c}/k-3)$}}}
\put(220.0,113.0){\rule[-0.200pt]{0.400pt}{71.065pt}}
\put(220,268){\circle*{16}}
\put(303,264){\circle{12}}
\put(350,263){\circle{12}}
\put(443,257){\circle{12}}
\put(622,248){\circle{12}}
\put(220.0,244.0){\rule[-0.200pt]{0.400pt}{11.563pt}}
\put(210.0,244.0){\rule[-0.200pt]{4.818pt}{0.400pt}}
\put(210.0,292.0){\rule[-0.200pt]{4.818pt}{0.400pt}}
\put(303.0,236.0){\rule[-0.200pt]{0.400pt}{13.490pt}}
\put(293.0,236.0){\rule[-0.200pt]{4.818pt}{0.400pt}}
\put(293.0,292.0){\rule[-0.200pt]{4.818pt}{0.400pt}}
\put(350.0,246.0){\rule[-0.200pt]{0.400pt}{8.431pt}}
\put(340.0,246.0){\rule[-0.200pt]{4.818pt}{0.400pt}}
\put(340.0,281.0){\rule[-0.200pt]{4.818pt}{0.400pt}}
\put(443.0,245.0){\rule[-0.200pt]{0.400pt}{5.541pt}}
\put(433.0,245.0){\rule[-0.200pt]{4.818pt}{0.400pt}}
\put(433.0,268.0){\rule[-0.200pt]{4.818pt}{0.400pt}}
\put(622.0,240.0){\rule[-0.200pt]{0.400pt}{4.095pt}}
\put(612.0,240.0){\rule[-0.200pt]{4.818pt}{0.400pt}}
\put(612.0,257.0){\rule[-0.200pt]{4.818pt}{0.400pt}}
\put(220,268){\usebox{\plotpoint}}
\put(237,266.67){\rule{1.445pt}{0.400pt}}
\multiput(237.00,267.17)(3.000,-1.000){2}{\rule{0.723pt}{0.400pt}}
\put(220.0,268.0){\rule[-0.200pt]{4.095pt}{0.400pt}}
\put(255,265.67){\rule{1.204pt}{0.400pt}}
\multiput(255.00,266.17)(2.500,-1.000){2}{\rule{0.602pt}{0.400pt}}
\put(243.0,267.0){\rule[-0.200pt]{2.891pt}{0.400pt}}
\put(278,264.67){\rule{1.204pt}{0.400pt}}
\multiput(278.00,265.17)(2.500,-1.000){2}{\rule{0.602pt}{0.400pt}}
\put(260.0,266.0){\rule[-0.200pt]{4.336pt}{0.400pt}}
\put(295,263.67){\rule{1.445pt}{0.400pt}}
\multiput(295.00,264.17)(3.000,-1.000){2}{\rule{0.723pt}{0.400pt}}
\put(283.0,265.0){\rule[-0.200pt]{2.891pt}{0.400pt}}
\put(318,262.67){\rule{1.445pt}{0.400pt}}
\multiput(318.00,263.17)(3.000,-1.000){2}{\rule{0.723pt}{0.400pt}}
\put(301.0,264.0){\rule[-0.200pt]{4.095pt}{0.400pt}}
\put(335,261.67){\rule{1.445pt}{0.400pt}}
\multiput(335.00,262.17)(3.000,-1.000){2}{\rule{0.723pt}{0.400pt}}
\put(324.0,263.0){\rule[-0.200pt]{2.650pt}{0.400pt}}
\put(358,260.67){\rule{1.445pt}{0.400pt}}
\multiput(358.00,261.17)(3.000,-1.000){2}{\rule{0.723pt}{0.400pt}}
\put(341.0,262.0){\rule[-0.200pt]{4.095pt}{0.400pt}}
\put(375,259.67){\rule{1.445pt}{0.400pt}}
\multiput(375.00,260.17)(3.000,-1.000){2}{\rule{0.723pt}{0.400pt}}
\put(364.0,261.0){\rule[-0.200pt]{2.650pt}{0.400pt}}
\put(398,258.67){\rule{1.445pt}{0.400pt}}
\multiput(398.00,259.17)(3.000,-1.000){2}{\rule{0.723pt}{0.400pt}}
\put(381.0,260.0){\rule[-0.200pt]{4.095pt}{0.400pt}}
\put(416,257.67){\rule{1.445pt}{0.400pt}}
\multiput(416.00,258.17)(3.000,-1.000){2}{\rule{0.723pt}{0.400pt}}
\put(404.0,259.0){\rule[-0.200pt]{2.891pt}{0.400pt}}
\put(433,256.67){\rule{1.445pt}{0.400pt}}
\multiput(433.00,257.17)(3.000,-1.000){2}{\rule{0.723pt}{0.400pt}}
\put(422.0,258.0){\rule[-0.200pt]{2.650pt}{0.400pt}}
\put(456,255.67){\rule{1.445pt}{0.400pt}}
\multiput(456.00,256.17)(3.000,-1.000){2}{\rule{0.723pt}{0.400pt}}
\put(439.0,257.0){\rule[-0.200pt]{4.095pt}{0.400pt}}
\put(473,254.67){\rule{1.445pt}{0.400pt}}
\multiput(473.00,255.17)(3.000,-1.000){2}{\rule{0.723pt}{0.400pt}}
\put(462.0,256.0){\rule[-0.200pt]{2.650pt}{0.400pt}}
\put(496,253.67){\rule{1.445pt}{0.400pt}}
\multiput(496.00,254.17)(3.000,-1.000){2}{\rule{0.723pt}{0.400pt}}
\put(479.0,255.0){\rule[-0.200pt]{4.095pt}{0.400pt}}
\put(514,252.67){\rule{1.204pt}{0.400pt}}
\multiput(514.00,253.17)(2.500,-1.000){2}{\rule{0.602pt}{0.400pt}}
\put(502.0,254.0){\rule[-0.200pt]{2.891pt}{0.400pt}}
\put(537,251.67){\rule{1.204pt}{0.400pt}}
\multiput(537.00,252.17)(2.500,-1.000){2}{\rule{0.602pt}{0.400pt}}
\put(519.0,253.0){\rule[-0.200pt]{4.336pt}{0.400pt}}
\put(554,250.67){\rule{1.445pt}{0.400pt}}
\multiput(554.00,251.17)(3.000,-1.000){2}{\rule{0.723pt}{0.400pt}}
\put(542.0,252.0){\rule[-0.200pt]{2.891pt}{0.400pt}}
\put(577,249.67){\rule{1.445pt}{0.400pt}}
\multiput(577.00,250.17)(3.000,-1.000){2}{\rule{0.723pt}{0.400pt}}
\put(560.0,251.0){\rule[-0.200pt]{4.095pt}{0.400pt}}
\put(594,248.67){\rule{1.445pt}{0.400pt}}
\multiput(594.00,249.17)(3.000,-1.000){2}{\rule{0.723pt}{0.400pt}}
\put(583.0,250.0){\rule[-0.200pt]{2.650pt}{0.400pt}}
\put(617,247.67){\rule{1.445pt}{0.400pt}}
\multiput(617.00,248.17)(3.000,-1.000){2}{\rule{0.723pt}{0.400pt}}
\put(600.0,249.0){\rule[-0.200pt]{4.095pt}{0.400pt}}
\put(635,246.67){\rule{1.204pt}{0.400pt}}
\multiput(635.00,247.17)(2.500,-1.000){2}{\rule{0.602pt}{0.400pt}}
\put(623.0,248.0){\rule[-0.200pt]{2.891pt}{0.400pt}}
\put(658,245.67){\rule{1.204pt}{0.400pt}}
\multiput(658.00,246.17)(2.500,-1.000){2}{\rule{0.602pt}{0.400pt}}
\put(640.0,247.0){\rule[-0.200pt]{4.336pt}{0.400pt}}
\put(675,244.67){\rule{1.445pt}{0.400pt}}
\multiput(675.00,245.17)(3.000,-1.000){2}{\rule{0.723pt}{0.400pt}}
\put(663.0,246.0){\rule[-0.200pt]{2.891pt}{0.400pt}}
\put(698,243.67){\rule{1.445pt}{0.400pt}}
\multiput(698.00,244.17)(3.000,-1.000){2}{\rule{0.723pt}{0.400pt}}
\put(681.0,245.0){\rule[-0.200pt]{4.095pt}{0.400pt}}
\put(715,242.67){\rule{1.445pt}{0.400pt}}
\multiput(715.00,243.17)(3.000,-1.000){2}{\rule{0.723pt}{0.400pt}}
\put(704.0,244.0){\rule[-0.200pt]{2.650pt}{0.400pt}}
\put(732,241.67){\rule{1.445pt}{0.400pt}}
\multiput(732.00,242.17)(3.000,-1.000){2}{\rule{0.723pt}{0.400pt}}
\put(721.0,243.0){\rule[-0.200pt]{2.650pt}{0.400pt}}
\put(755,240.67){\rule{1.445pt}{0.400pt}}
\multiput(755.00,241.17)(3.000,-1.000){2}{\rule{0.723pt}{0.400pt}}
\put(738.0,242.0){\rule[-0.200pt]{4.095pt}{0.400pt}}
\put(773,239.67){\rule{1.204pt}{0.400pt}}
\multiput(773.00,240.17)(2.500,-1.000){2}{\rule{0.602pt}{0.400pt}}
\put(761.0,241.0){\rule[-0.200pt]{2.891pt}{0.400pt}}
\put(778.0,240.0){\rule[-0.200pt]{2.891pt}{0.400pt}}
\end{picture}
\]
\vspace{-0.2in}
\caption{Chiral extrapolation for $J_{q, CI}$ as a function of the 
quark mass. The value at the chiral limit is indicated by $\bullet$.}
\end{figure}
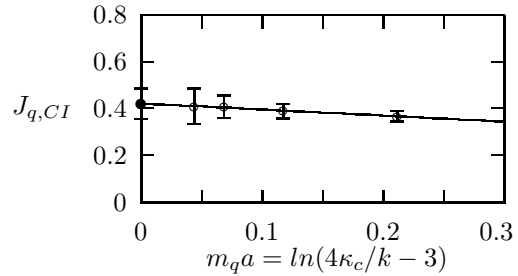
For the DI contribution, we follow the calculations for the flavor-singlet
axial coupling~\cite{dll95}, the $\pi N \sigma$ term~\cite{dll96}, and
the strangeness magnetic moment~\cite{dlw98} by summing the current insertion
time $t_1$ from the nucleon source to its sink in the corresponding
ratio in Eq. (\ref{ratio}) for the DI (Fig. 1(b)) to gain statistics. In 
this case, the ratio leads to $const + 1/2 \epsilon_{jkm} q_k T_{DI}(q^2)
\, t_2$. We take the average of the 3 polarization directions. The
DI result is calculated from the slope of this summed ratio 
with respect to $t_2$. To evaluate the trace of the quark loop in Fig. 1(b), 
we adopt the same stochastic algorithm with the $Z_2$ noise estimator~\cite
{dl94} as in other DI calculations~\cite{dll95,dll96,dlw98}. 
In addition, we shall use two more techniques to reduce the errors from the
stochastic algorithm. First one is to observe that from the charge 
conjugation and Euclidean hermiticity (CH symmetry), the three-point 
function for the DI 
which is the product of the quark loop and the nucleon propagator is real. 
On the other hand, the Euclidean hermiticy itself dictates that the loop 
should also be real. Therefore, we need only multiply the real part of the
loop with the real part of the nucleon propagator and neglect the product of 
the imaginary parts of them which only introduces noise to 
the signal. Secondly, we employ an unbiased subtraction method
which has been used in the calculation of the fermion 
determinant~\cite{tdl98}. 
The trace of the inverse matrix $A^{-1}$ can be estimated stochastically
as follows 
\begin{equation}
\hbox{Tr}( A^{-1} )
= E [ < \eta^{\dagger} ( A^{-1} 
- \sum^{P}_{i=1} \lambda_{i}\, O^{(i)} ) \eta > ],
\end{equation}
where $\eta$'s are $Z_{2}$ noise vectors, $O^{(i)}$'s are a set of $P$ 
traceless matrices and $\lambda_{i}$'s are the variational parameters which 
are determined by reducing the variance of the three-point function over
the gauge configurations. 
 In practice, we found that 
a judicious choice of $O^{(i)}$ is a set of traceless matrices from the 
hoping expansion 
of the propagator. Since they match the off-diagonal behavior of the matrix 
$A^{-1}$, they can offset the off-diagonal contribution to the 
variance~\cite{tdl98}. This 
method proves to be very efficient in reducing the error of the DI 
calculation with negligible overhead. After implementing the CH and 
H symmetries and the unbiased 
subtraction with traceless matrices obtained from just the first two terms of 
the hopping expansion, we obtain a reduction of error of 3 -- 4 times. 
Fig.4 shows a plot of the summed three-point to two-point function ratio
(as in Eq. (\ref{ratio}) for the DI) {\it vs.} $t_2$ for 
$\kappa = 0.154$ and $|\vec{q}| = 2\pi/La$ with and without subtraction. 
One can see that the error bars before subtraction are 
much larger and only with the subtraction does one get a reasonably good
slope as illustrated by the fitted straight line.  
With the help of this subtraction procedure we calculate slopes for 
other $q^{2}$ and for other $\kappa's$ which are shown in Fig.5. We use fixed 
source and vary the sink position $t_2$. From $t_2 = 8$ on, the nucleon 
becomes isolated from its excited states~\cite{ldd94}. Hence, the slopes 
are fitted in the region $t_2 \geq 8$ (Fig. 5). 
 Next, $T_{DI}(q^2)$ is fitted 
with a monopole form in $q^2$ as in the other DI 
calculations~\cite{dll96,dlw98}.
These are plotted in Fig. 6. Similar to the CI case, we also use covariant 
matrix fitting and the final error bars are obtained by the jackknife method. 
A finite mass correction factor from the triangle diagram~\cite{ll95} is 
introduced while extrapolating to the chiral limit with a linear 
$m_q a$ dependence. This is shown in Fig. 7(a). The strange quark 
contribution is obtained by fixing the sea quark mass at $\kappa_s = 0.154$ and
extrapolate the valence $\kappa_v$ from $0.148, 0.152, 0.154$ to $\kappa_c$.
This is shown in Fig. 7(b). It is interesting to point out that the
result is fairly independent of the sea quark mass in the fermion loop (
Fig. 1(b)). Comparing Fig. 7(a) where the valence and sea quarks are kept the
same ($\kappa_v = \kappa_s$) and Fig. 7(b), we see that albeit the 
sea quark mass in Fig. 7(a)
changes by a factor of 3, the result still coincide with those in Fig. 7(b)
for each of the valence-quark case. This shows that the DI depends on the
valence-quark mass but is almost independent of the sea-quark mass. 

\vspace*{-0.15in}
\begin{figure}[h]
\[
\input{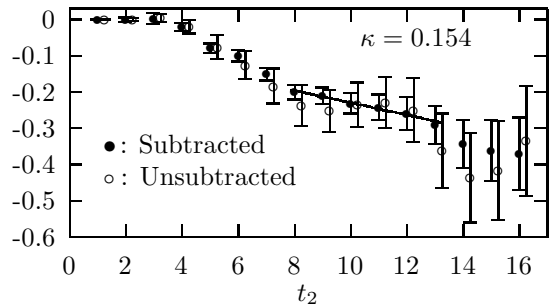}
\]
\vspace*{-0.25in}
\caption{The summed ratios of Eq.(\ref{ratio}) for the DI are plotted for 
different time slice $t_2$ with and without unbiased subtraction. Ratios 
without subtraction are shifted slightly towards right.}
\end{figure}
\vspace*{-0.25in}
\begin{figure}
\[
\hspace*{-0.4in}
\setlength{\unitlength}{0.240900pt}
\ifx\plotpoint\undefined\newsavebox{\plotpoint}\fi
\sbox{\plotpoint}{\rule[-0.200pt]{0.400pt}{0.400pt}}%
\begin{picture}(510,467)(0,0)
\font\gnuplot=cmr10 at 10pt
\gnuplot
\sbox{\plotpoint}{\rule[-0.200pt]{0.400pt}{0.400pt}}%
\put(176.0,113.0){\rule[-0.200pt]{4.818pt}{0.400pt}}
\put(170,113){\makebox(0,0)[r]{-0.3}}
\put(426.0,113.0){\rule[-0.200pt]{4.818pt}{0.400pt}}
\put(176.0,195.0){\rule[-0.200pt]{4.818pt}{0.400pt}}
\put(170,195){\makebox(0,0)[r]{-0.2}}
\put(426.0,195.0){\rule[-0.200pt]{4.818pt}{0.400pt}}
\put(176.0,276.0){\rule[-0.200pt]{4.818pt}{0.400pt}}
\put(170,276){\makebox(0,0)[r]{-0.1}}
\put(426.0,276.0){\rule[-0.200pt]{4.818pt}{0.400pt}}
\put(176.0,358.0){\rule[-0.200pt]{4.818pt}{0.400pt}}
\put(170,358){\makebox(0,0)[r]{0}}
\put(426.0,358.0){\rule[-0.200pt]{4.818pt}{0.400pt}}
\put(176.0,113.0){\rule[-0.200pt]{0.400pt}{4.818pt}}
\put(176,68){\makebox(0,0){0}}
\put(176.0,379.0){\rule[-0.200pt]{0.400pt}{4.818pt}}
\put(240.0,113.0){\rule[-0.200pt]{0.400pt}{4.818pt}}
\put(240,68){\makebox(0,0){4}}
\put(240.0,379.0){\rule[-0.200pt]{0.400pt}{4.818pt}}
\put(303.0,113.0){\rule[-0.200pt]{0.400pt}{4.818pt}}
\put(303,68){\makebox(0,0){8}}
\put(303.0,379.0){\rule[-0.200pt]{0.400pt}{4.818pt}}
\put(367.0,113.0){\rule[-0.200pt]{0.400pt}{4.818pt}}
\put(367,68){\makebox(0,0){12}}
\put(367.0,379.0){\rule[-0.200pt]{0.400pt}{4.818pt}}
\put(430.0,113.0){\rule[-0.200pt]{0.400pt}{4.818pt}}
\put(430,68){\makebox(0,0){16}}
\put(430.0,379.0){\rule[-0.200pt]{0.400pt}{4.818pt}}
\put(176.0,113.0){\rule[-0.200pt]{65.043pt}{0.400pt}}
\put(446.0,113.0){\rule[-0.200pt]{0.400pt}{68.897pt}}
\put(176.0,399.0){\rule[-0.200pt]{65.043pt}{0.400pt}}
\put(311,23){\makebox(0,0){{$t_2$}}}
\put(300,180){\makebox(0,0){{$\kappa = 0.148$}}}
\put(176.0,113.0){\rule[-0.200pt]{0.400pt}{68.897pt}}
\put(192,358){\circle{12}}
\put(208,357){\circle{12}}
\put(224,356){\circle{12}}
\put(240,347){\circle{12}}
\put(255,337){\circle{12}}
\put(271,325){\circle{12}}
\put(287,312){\circle{12}}
\put(303,302){\circle{12}}
\put(319,294){\circle{12}}
\put(335,293){\circle{12}}
\put(351,292){\circle{12}}
\put(367,287){\circle{12}}
\put(382,279){\circle{12}}
\put(398,269){\circle{12}}
\put(414,262){\circle{12}}
\put(430,263){\circle{12}}
\put(192,358){\usebox{\plotpoint}}
\put(182.0,358.0){\rule[-0.200pt]{4.818pt}{0.400pt}}
\put(182.0,358.0){\rule[-0.200pt]{4.818pt}{0.400pt}}
\put(208.0,356.0){\usebox{\plotpoint}}
\put(198.0,356.0){\rule[-0.200pt]{4.818pt}{0.400pt}}
\put(198.0,357.0){\rule[-0.200pt]{4.818pt}{0.400pt}}
\put(224.0,354.0){\rule[-0.200pt]{0.400pt}{0.723pt}}
\put(214.0,354.0){\rule[-0.200pt]{4.818pt}{0.400pt}}
\put(214.0,357.0){\rule[-0.200pt]{4.818pt}{0.400pt}}
\put(240.0,344.0){\rule[-0.200pt]{0.400pt}{1.445pt}}
\put(230.0,344.0){\rule[-0.200pt]{4.818pt}{0.400pt}}
\put(230.0,350.0){\rule[-0.200pt]{4.818pt}{0.400pt}}
\put(255.0,334.0){\rule[-0.200pt]{0.400pt}{1.686pt}}
\put(245.0,334.0){\rule[-0.200pt]{4.818pt}{0.400pt}}
\put(245.0,341.0){\rule[-0.200pt]{4.818pt}{0.400pt}}
\put(271.0,321.0){\rule[-0.200pt]{0.400pt}{1.927pt}}
\put(261.0,321.0){\rule[-0.200pt]{4.818pt}{0.400pt}}
\put(261.0,329.0){\rule[-0.200pt]{4.818pt}{0.400pt}}
\put(287.0,307.0){\rule[-0.200pt]{0.400pt}{2.409pt}}
\put(277.0,307.0){\rule[-0.200pt]{4.818pt}{0.400pt}}
\put(277.0,317.0){\rule[-0.200pt]{4.818pt}{0.400pt}}
\put(303.0,297.0){\rule[-0.200pt]{0.400pt}{2.409pt}}
\put(293.0,297.0){\rule[-0.200pt]{4.818pt}{0.400pt}}
\put(293.0,307.0){\rule[-0.200pt]{4.818pt}{0.400pt}}
\put(319.0,288.0){\rule[-0.200pt]{0.400pt}{2.891pt}}
\put(309.0,288.0){\rule[-0.200pt]{4.818pt}{0.400pt}}
\put(309.0,300.0){\rule[-0.200pt]{4.818pt}{0.400pt}}
\put(335.0,286.0){\rule[-0.200pt]{0.400pt}{3.373pt}}
\put(325.0,286.0){\rule[-0.200pt]{4.818pt}{0.400pt}}
\put(325.0,300.0){\rule[-0.200pt]{4.818pt}{0.400pt}}
\put(351.0,283.0){\rule[-0.200pt]{0.400pt}{4.095pt}}
\put(341.0,283.0){\rule[-0.200pt]{4.818pt}{0.400pt}}
\put(341.0,300.0){\rule[-0.200pt]{4.818pt}{0.400pt}}
\put(367.0,276.0){\rule[-0.200pt]{0.400pt}{5.059pt}}
\put(357.0,276.0){\rule[-0.200pt]{4.818pt}{0.400pt}}
\put(357.0,297.0){\rule[-0.200pt]{4.818pt}{0.400pt}}
\put(382.0,268.0){\rule[-0.200pt]{0.400pt}{5.541pt}}
\put(372.0,268.0){\rule[-0.200pt]{4.818pt}{0.400pt}}
\put(372.0,291.0){\rule[-0.200pt]{4.818pt}{0.400pt}}
\put(398.0,256.0){\rule[-0.200pt]{0.400pt}{6.263pt}}
\put(388.0,256.0){\rule[-0.200pt]{4.818pt}{0.400pt}}
\put(388.0,282.0){\rule[-0.200pt]{4.818pt}{0.400pt}}
\put(414.0,247.0){\rule[-0.200pt]{0.400pt}{6.986pt}}
\put(404.0,247.0){\rule[-0.200pt]{4.818pt}{0.400pt}}
\put(404.0,276.0){\rule[-0.200pt]{4.818pt}{0.400pt}}
\put(430.0,247.0){\rule[-0.200pt]{0.400pt}{7.709pt}}
\put(420.0,247.0){\rule[-0.200pt]{4.818pt}{0.400pt}}
\put(420.0,279.0){\rule[-0.200pt]{4.818pt}{0.400pt}}
\put(176,373){\usebox{\plotpoint}}
\put(329,301.17){\rule{0.482pt}{0.400pt}}
\multiput(329.00,302.17)(1.000,-2.000){2}{\rule{0.241pt}{0.400pt}}
\put(331,299.67){\rule{0.723pt}{0.400pt}}
\multiput(331.00,300.17)(1.500,-1.000){2}{\rule{0.361pt}{0.400pt}}
\put(334,298.67){\rule{0.723pt}{0.400pt}}
\multiput(334.00,299.17)(1.500,-1.000){2}{\rule{0.361pt}{0.400pt}}
\put(337,297.67){\rule{0.723pt}{0.400pt}}
\multiput(337.00,298.17)(1.500,-1.000){2}{\rule{0.361pt}{0.400pt}}
\put(340,296.17){\rule{0.482pt}{0.400pt}}
\multiput(340.00,297.17)(1.000,-2.000){2}{\rule{0.241pt}{0.400pt}}
\put(342,294.67){\rule{0.723pt}{0.400pt}}
\multiput(342.00,295.17)(1.500,-1.000){2}{\rule{0.361pt}{0.400pt}}
\put(345,293.67){\rule{0.723pt}{0.400pt}}
\multiput(345.00,294.17)(1.500,-1.000){2}{\rule{0.361pt}{0.400pt}}
\put(348,292.67){\rule{0.723pt}{0.400pt}}
\multiput(348.00,293.17)(1.500,-1.000){2}{\rule{0.361pt}{0.400pt}}
\put(351,291.17){\rule{0.482pt}{0.400pt}}
\multiput(351.00,292.17)(1.000,-2.000){2}{\rule{0.241pt}{0.400pt}}
\put(353,289.67){\rule{0.723pt}{0.400pt}}
\multiput(353.00,290.17)(1.500,-1.000){2}{\rule{0.361pt}{0.400pt}}
\put(356,288.67){\rule{0.723pt}{0.400pt}}
\multiput(356.00,289.17)(1.500,-1.000){2}{\rule{0.361pt}{0.400pt}}
\put(359,287.67){\rule{0.482pt}{0.400pt}}
\multiput(359.00,288.17)(1.000,-1.000){2}{\rule{0.241pt}{0.400pt}}
\put(361,286.17){\rule{0.700pt}{0.400pt}}
\multiput(361.00,287.17)(1.547,-2.000){2}{\rule{0.350pt}{0.400pt}}
\put(364,284.67){\rule{0.723pt}{0.400pt}}
\multiput(364.00,285.17)(1.500,-1.000){2}{\rule{0.361pt}{0.400pt}}
\put(367,283.67){\rule{0.723pt}{0.400pt}}
\multiput(367.00,284.17)(1.500,-1.000){2}{\rule{0.361pt}{0.400pt}}
\put(370,282.67){\rule{0.482pt}{0.400pt}}
\multiput(370.00,283.17)(1.000,-1.000){2}{\rule{0.241pt}{0.400pt}}
\put(372,281.17){\rule{0.700pt}{0.400pt}}
\multiput(372.00,282.17)(1.547,-2.000){2}{\rule{0.350pt}{0.400pt}}
\put(375,279.67){\rule{0.723pt}{0.400pt}}
\multiput(375.00,280.17)(1.500,-1.000){2}{\rule{0.361pt}{0.400pt}}
\put(378,278.67){\rule{0.723pt}{0.400pt}}
\multiput(378.00,279.17)(1.500,-1.000){2}{\rule{0.361pt}{0.400pt}}
\put(381,277.67){\rule{0.482pt}{0.400pt}}
\multiput(381.00,278.17)(1.000,-1.000){2}{\rule{0.241pt}{0.400pt}}
\put(383,276.17){\rule{0.700pt}{0.400pt}}
\multiput(383.00,277.17)(1.547,-2.000){2}{\rule{0.350pt}{0.400pt}}
\put(386,274.67){\rule{0.723pt}{0.400pt}}
\multiput(386.00,275.17)(1.500,-1.000){2}{\rule{0.361pt}{0.400pt}}
\put(389,273.67){\rule{0.482pt}{0.400pt}}
\multiput(389.00,274.17)(1.000,-1.000){2}{\rule{0.241pt}{0.400pt}}
\put(391,272.67){\rule{0.723pt}{0.400pt}}
\multiput(391.00,273.17)(1.500,-1.000){2}{\rule{0.361pt}{0.400pt}}
\put(394,271.17){\rule{0.700pt}{0.400pt}}
\multiput(394.00,272.17)(1.547,-2.000){2}{\rule{0.350pt}{0.400pt}}
\put(397,269.67){\rule{0.723pt}{0.400pt}}
\multiput(397.00,270.17)(1.500,-1.000){2}{\rule{0.361pt}{0.400pt}}
\put(400,268.67){\rule{0.482pt}{0.400pt}}
\multiput(400.00,269.17)(1.000,-1.000){2}{\rule{0.241pt}{0.400pt}}
\put(402,267.67){\rule{0.723pt}{0.400pt}}
\multiput(402.00,268.17)(1.500,-1.000){2}{\rule{0.361pt}{0.400pt}}
\end{picture}
\hspace*{-0.5in}
\setlength{\unitlength}{0.240900pt}
\ifx\plotpoint\undefined\newsavebox{\plotpoint}\fi
\sbox{\plotpoint}{\rule[-0.200pt]{0.400pt}{0.400pt}}%
\begin{picture}(510,467)(0,0)
\font\gnuplot=cmr10 at 10pt
\gnuplot
\sbox{\plotpoint}{\rule[-0.200pt]{0.400pt}{0.400pt}}%
\put(176.0,113.0){\rule[-0.200pt]{4.818pt}{0.400pt}}
\put(170,113){\makebox(0,0)[r]{-0.4}}
\put(426.0,113.0){\rule[-0.200pt]{4.818pt}{0.400pt}}
\put(176.0,177.0){\rule[-0.200pt]{4.818pt}{0.400pt}}
\put(170,177){\makebox(0,0)[r]{-0.3}}
\put(426.0,177.0){\rule[-0.200pt]{4.818pt}{0.400pt}}
\put(176.0,240.0){\rule[-0.200pt]{4.818pt}{0.400pt}}
\put(170,240){\makebox(0,0)[r]{-0.2}}
\put(426.0,240.0){\rule[-0.200pt]{4.818pt}{0.400pt}}
\put(176.0,304.0){\rule[-0.200pt]{4.818pt}{0.400pt}}
\put(170,304){\makebox(0,0)[r]{-0.1}}
\put(426.0,304.0){\rule[-0.200pt]{4.818pt}{0.400pt}}
\put(176.0,367.0){\rule[-0.200pt]{4.818pt}{0.400pt}}
\put(170,367){\makebox(0,0)[r]{0}}
\put(426.0,367.0){\rule[-0.200pt]{4.818pt}{0.400pt}}
\put(176.0,113.0){\rule[-0.200pt]{0.400pt}{4.818pt}}
\put(176,68){\makebox(0,0){0}}
\put(176.0,379.0){\rule[-0.200pt]{0.400pt}{4.818pt}}
\put(240.0,113.0){\rule[-0.200pt]{0.400pt}{4.818pt}}
\put(240,68){\makebox(0,0){4}}
\put(240.0,379.0){\rule[-0.200pt]{0.400pt}{4.818pt}}
\put(303.0,113.0){\rule[-0.200pt]{0.400pt}{4.818pt}}
\put(303,68){\makebox(0,0){8}}
\put(303.0,379.0){\rule[-0.200pt]{0.400pt}{4.818pt}}
\put(367.0,113.0){\rule[-0.200pt]{0.400pt}{4.818pt}}
\put(367,68){\makebox(0,0){12}}
\put(367.0,379.0){\rule[-0.200pt]{0.400pt}{4.818pt}}
\put(430.0,113.0){\rule[-0.200pt]{0.400pt}{4.818pt}}
\put(430,68){\makebox(0,0){16}}
\put(430.0,379.0){\rule[-0.200pt]{0.400pt}{4.818pt}}
\put(176.0,113.0){\rule[-0.200pt]{65.043pt}{0.400pt}}
\put(446.0,113.0){\rule[-0.200pt]{0.400pt}{68.897pt}}
\put(176.0,399.0){\rule[-0.200pt]{65.043pt}{0.400pt}}
\put(311,23){\makebox(0,0){{$t_2$}}}
\put(300,180){\makebox(0,0){{$\kappa = 0.152$}}}
\put(176.0,113.0){\rule[-0.200pt]{0.400pt}{68.897pt}}
\put(192,367){\circle{12}}
\put(208,366){\circle{12}}
\put(224,367){\circle{12}}
\put(240,354){\circle{12}}
\put(255,338){\circle{12}}
\put(271,319){\circle{12}}
\put(287,310){\circle{12}}
\put(303,288){\circle{12}}
\put(319,283){\circle{12}}
\put(335,281){\circle{12}}
\put(351,276){\circle{12}}
\put(367,263){\circle{12}}
\put(382,253){\circle{12}}
\put(398,241){\circle{12}}
\put(414,232){\circle{12}}
\put(430,224){\circle{12}}
\put(192,367){\usebox{\plotpoint}}
\put(182.0,367.0){\rule[-0.200pt]{4.818pt}{0.400pt}}
\put(182.0,367.0){\rule[-0.200pt]{4.818pt}{0.400pt}}
\put(208.0,364.0){\rule[-0.200pt]{0.400pt}{1.204pt}}
\put(198.0,364.0){\rule[-0.200pt]{4.818pt}{0.400pt}}
\put(198.0,369.0){\rule[-0.200pt]{4.818pt}{0.400pt}}
\put(224.0,361.0){\rule[-0.200pt]{0.400pt}{2.650pt}}
\put(214.0,361.0){\rule[-0.200pt]{4.818pt}{0.400pt}}
\put(214.0,372.0){\rule[-0.200pt]{4.818pt}{0.400pt}}
\put(240.0,349.0){\rule[-0.200pt]{0.400pt}{2.409pt}}
\put(230.0,349.0){\rule[-0.200pt]{4.818pt}{0.400pt}}
\put(230.0,359.0){\rule[-0.200pt]{4.818pt}{0.400pt}}
\put(255.0,331.0){\rule[-0.200pt]{0.400pt}{3.373pt}}
\put(245.0,331.0){\rule[-0.200pt]{4.818pt}{0.400pt}}
\put(245.0,345.0){\rule[-0.200pt]{4.818pt}{0.400pt}}
\put(271.0,312.0){\rule[-0.200pt]{0.400pt}{3.373pt}}
\put(261.0,312.0){\rule[-0.200pt]{4.818pt}{0.400pt}}
\put(261.0,326.0){\rule[-0.200pt]{4.818pt}{0.400pt}}
\put(287.0,304.0){\rule[-0.200pt]{0.400pt}{3.132pt}}
\put(277.0,304.0){\rule[-0.200pt]{4.818pt}{0.400pt}}
\put(277.0,317.0){\rule[-0.200pt]{4.818pt}{0.400pt}}
\put(303.0,281.0){\rule[-0.200pt]{0.400pt}{3.613pt}}
\put(293.0,281.0){\rule[-0.200pt]{4.818pt}{0.400pt}}
\put(293.0,296.0){\rule[-0.200pt]{4.818pt}{0.400pt}}
\put(319.0,275.0){\rule[-0.200pt]{0.400pt}{4.095pt}}
\put(309.0,275.0){\rule[-0.200pt]{4.818pt}{0.400pt}}
\put(309.0,292.0){\rule[-0.200pt]{4.818pt}{0.400pt}}
\put(335.0,272.0){\rule[-0.200pt]{0.400pt}{4.336pt}}
\put(325.0,272.0){\rule[-0.200pt]{4.818pt}{0.400pt}}
\put(325.0,290.0){\rule[-0.200pt]{4.818pt}{0.400pt}}
\put(351.0,266.0){\rule[-0.200pt]{0.400pt}{4.818pt}}
\put(341.0,266.0){\rule[-0.200pt]{4.818pt}{0.400pt}}
\put(341.0,286.0){\rule[-0.200pt]{4.818pt}{0.400pt}}
\put(367.0,253.0){\rule[-0.200pt]{0.400pt}{4.818pt}}
\put(357.0,253.0){\rule[-0.200pt]{4.818pt}{0.400pt}}
\put(357.0,273.0){\rule[-0.200pt]{4.818pt}{0.400pt}}
\put(382.0,241.0){\rule[-0.200pt]{0.400pt}{5.782pt}}
\put(372.0,241.0){\rule[-0.200pt]{4.818pt}{0.400pt}}
\put(372.0,265.0){\rule[-0.200pt]{4.818pt}{0.400pt}}
\put(398.0,228.0){\rule[-0.200pt]{0.400pt}{6.263pt}}
\put(388.0,228.0){\rule[-0.200pt]{4.818pt}{0.400pt}}
\put(388.0,254.0){\rule[-0.200pt]{4.818pt}{0.400pt}}
\put(414.0,211.0){\rule[-0.200pt]{0.400pt}{10.359pt}}
\put(404.0,211.0){\rule[-0.200pt]{4.818pt}{0.400pt}}
\put(404.0,254.0){\rule[-0.200pt]{4.818pt}{0.400pt}}
\put(430.0,186.0){\rule[-0.200pt]{0.400pt}{18.549pt}}
\put(420.0,186.0){\rule[-0.200pt]{4.818pt}{0.400pt}}
\put(420.0,263.0){\rule[-0.200pt]{4.818pt}{0.400pt}}
\put(176,382){\usebox{\plotpoint}}
\put(321,291.17){\rule{0.482pt}{0.400pt}}
\multiput(321.00,292.17)(1.000,-2.000){2}{\rule{0.241pt}{0.400pt}}
\put(323,289.17){\rule{0.700pt}{0.400pt}}
\multiput(323.00,290.17)(1.547,-2.000){2}{\rule{0.350pt}{0.400pt}}
\put(326,287.67){\rule{0.723pt}{0.400pt}}
\multiput(326.00,288.17)(1.500,-1.000){2}{\rule{0.361pt}{0.400pt}}
\put(329,286.17){\rule{0.482pt}{0.400pt}}
\multiput(329.00,287.17)(1.000,-2.000){2}{\rule{0.241pt}{0.400pt}}
\put(331,284.17){\rule{0.700pt}{0.400pt}}
\multiput(331.00,285.17)(1.547,-2.000){2}{\rule{0.350pt}{0.400pt}}
\put(334,282.17){\rule{0.700pt}{0.400pt}}
\multiput(334.00,283.17)(1.547,-2.000){2}{\rule{0.350pt}{0.400pt}}
\put(337,280.67){\rule{0.723pt}{0.400pt}}
\multiput(337.00,281.17)(1.500,-1.000){2}{\rule{0.361pt}{0.400pt}}
\put(340,279.17){\rule{0.482pt}{0.400pt}}
\multiput(340.00,280.17)(1.000,-2.000){2}{\rule{0.241pt}{0.400pt}}
\put(342,277.17){\rule{0.700pt}{0.400pt}}
\multiput(342.00,278.17)(1.547,-2.000){2}{\rule{0.350pt}{0.400pt}}
\put(345,275.67){\rule{0.723pt}{0.400pt}}
\multiput(345.00,276.17)(1.500,-1.000){2}{\rule{0.361pt}{0.400pt}}
\put(348,274.17){\rule{0.700pt}{0.400pt}}
\multiput(348.00,275.17)(1.547,-2.000){2}{\rule{0.350pt}{0.400pt}}
\put(351,272.17){\rule{0.482pt}{0.400pt}}
\multiput(351.00,273.17)(1.000,-2.000){2}{\rule{0.241pt}{0.400pt}}
\put(353,270.67){\rule{0.723pt}{0.400pt}}
\multiput(353.00,271.17)(1.500,-1.000){2}{\rule{0.361pt}{0.400pt}}
\put(356,269.17){\rule{0.700pt}{0.400pt}}
\multiput(356.00,270.17)(1.547,-2.000){2}{\rule{0.350pt}{0.400pt}}
\put(359,267.17){\rule{0.482pt}{0.400pt}}
\multiput(359.00,268.17)(1.000,-2.000){2}{\rule{0.241pt}{0.400pt}}
\put(361,265.67){\rule{0.723pt}{0.400pt}}
\multiput(361.00,266.17)(1.500,-1.000){2}{\rule{0.361pt}{0.400pt}}
\put(364,264.17){\rule{0.700pt}{0.400pt}}
\multiput(364.00,265.17)(1.547,-2.000){2}{\rule{0.350pt}{0.400pt}}
\put(367,262.17){\rule{0.700pt}{0.400pt}}
\multiput(367.00,263.17)(1.547,-2.000){2}{\rule{0.350pt}{0.400pt}}
\put(370,260.67){\rule{0.482pt}{0.400pt}}
\multiput(370.00,261.17)(1.000,-1.000){2}{\rule{0.241pt}{0.400pt}}
\put(372,259.17){\rule{0.700pt}{0.400pt}}
\multiput(372.00,260.17)(1.547,-2.000){2}{\rule{0.350pt}{0.400pt}}
\put(375,257.17){\rule{0.700pt}{0.400pt}}
\multiput(375.00,258.17)(1.547,-2.000){2}{\rule{0.350pt}{0.400pt}}
\put(378,255.17){\rule{0.700pt}{0.400pt}}
\multiput(378.00,256.17)(1.547,-2.000){2}{\rule{0.350pt}{0.400pt}}
\put(381,253.67){\rule{0.482pt}{0.400pt}}
\multiput(381.00,254.17)(1.000,-1.000){2}{\rule{0.241pt}{0.400pt}}
\put(383,252.17){\rule{0.700pt}{0.400pt}}
\multiput(383.00,253.17)(1.547,-2.000){2}{\rule{0.350pt}{0.400pt}}
\put(386,250.17){\rule{0.700pt}{0.400pt}}
\multiput(386.00,251.17)(1.547,-2.000){2}{\rule{0.350pt}{0.400pt}}
\put(389,248.67){\rule{0.482pt}{0.400pt}}
\multiput(389.00,249.17)(1.000,-1.000){2}{\rule{0.241pt}{0.400pt}}
\put(391,247.17){\rule{0.700pt}{0.400pt}}
\multiput(391.00,248.17)(1.547,-2.000){2}{\rule{0.350pt}{0.400pt}}
\put(394,245.17){\rule{0.700pt}{0.400pt}}
\multiput(394.00,246.17)(1.547,-2.000){2}{\rule{0.350pt}{0.400pt}}
\put(397,243.67){\rule{0.723pt}{0.400pt}}
\multiput(397.00,244.17)(1.500,-1.000){2}{\rule{0.361pt}{0.400pt}}
\put(400,242.17){\rule{0.482pt}{0.400pt}}
\multiput(400.00,243.17)(1.000,-2.000){2}{\rule{0.241pt}{0.400pt}}
\put(402,240.17){\rule{0.700pt}{0.400pt}}
\multiput(402.00,241.17)(1.547,-2.000){2}{\rule{0.350pt}{0.400pt}}
\end{picture}
\hspace*{-0.5in}
\input{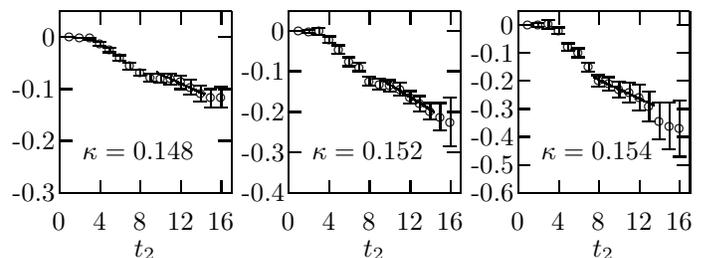}
\]
\vspace*{-0.25in}
\caption{The summed ratios of Eq.(\ref{ratio}) for the DI are plotted for 
 $|\vec{q}| = 2\pi/La$.}
\end{figure}
\vspace*{-0.2in}
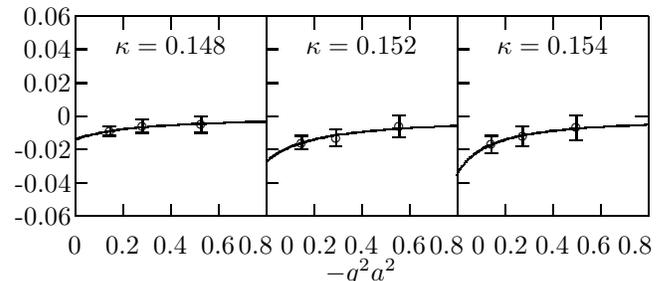
\begin{figure}[h]
\[
\hspace*{-0.25in}
\setlength{\unitlength}{0.240900pt}
\ifx\plotpoint\undefined\newsavebox{\plotpoint}\fi
\sbox{\plotpoint}{\rule[-0.200pt]{0.400pt}{0.400pt}}%
\begin{picture}(540,450)(0,0)
\font\gnuplot=cmr10 at 10pt
\gnuplot
\sbox{\plotpoint}{\rule[-0.200pt]{0.400pt}{0.400pt}}%
\put(176.0,68.0){\rule[-0.200pt]{4.818pt}{0.400pt}}
\put(170,68){\makebox(0,0)[r]{-0.06}}
\put(456.0,68.0){\rule[-0.200pt]{4.818pt}{0.400pt}}
\put(176.0,120.0){\rule[-0.200pt]{4.818pt}{0.400pt}}
\put(170,120){\makebox(0,0)[r]{-0.04}}
\put(176.0,173.0){\rule[-0.200pt]{4.818pt}{0.400pt}}
\put(170,173){\makebox(0,0)[r]{-0.02}}
\put(176.0,225.0){\rule[-0.200pt]{4.818pt}{0.400pt}}
\put(170,225){\makebox(0,0)[r]{0}}
\put(176.0,277.0){\rule[-0.200pt]{4.818pt}{0.400pt}}
\put(170,277){\makebox(0,0)[r]{0.02}}
\put(176.0,330.0){\rule[-0.200pt]{4.818pt}{0.400pt}}
\put(170,330){\makebox(0,0)[r]{0.04}}
\put(176.0,382.0){\rule[-0.200pt]{4.818pt}{0.400pt}}
\put(170,382){\makebox(0,0)[r]{0.06}}
\put(456.0,382.0){\rule[-0.200pt]{4.818pt}{0.400pt}}
\put(176.0,68.0){\rule[-0.200pt]{0.400pt}{4.818pt}}
\put(176,23){\makebox(0,0){0}}
\put(176.0,362.0){\rule[-0.200pt]{0.400pt}{4.818pt}}
\put(251.0,68.0){\rule[-0.200pt]{0.400pt}{4.818pt}}
\put(251,23){\makebox(0,0){0.2}}
\put(251.0,362.0){\rule[-0.200pt]{0.400pt}{4.818pt}}
\put(326.0,68.0){\rule[-0.200pt]{0.400pt}{4.818pt}}
\put(326,23){\makebox(0,0){0.4}}
\put(326.0,362.0){\rule[-0.200pt]{0.400pt}{4.818pt}}
\put(401.0,68.0){\rule[-0.200pt]{0.400pt}{4.818pt}}
\put(401,23){\makebox(0,0){0.6}}
\put(401.0,362.0){\rule[-0.200pt]{0.400pt}{4.818pt}}
\put(476.0,68.0){\rule[-0.200pt]{0.400pt}{4.818pt}}
\put(460,23){\makebox(0,0){0.8}}
\put(476.0,362.0){\rule[-0.200pt]{0.400pt}{4.818pt}}
\put(176.0,68.0){\rule[-0.200pt]{72.270pt}{0.400pt}}
\put(476.0,68.0){\rule[-0.200pt]{0.400pt}{75.643pt}}
\put(176.0,382.0){\rule[-0.200pt]{72.270pt}{0.400pt}}
\put(326,337){\makebox(0,0){{$\kappa = 0.148$}}}
\put(176.0,68.0){\rule[-0.200pt]{0.400pt}{75.643pt}}
\put(231,201){\circle{12}}
\put(282,209){\circle{12}}
\put(374,212){\circle{12}}
\put(231.0,194.0){\rule[-0.200pt]{0.400pt}{3.613pt}}
\put(221.0,194.0){\rule[-0.200pt]{4.818pt}{0.400pt}}
\put(221.0,209.0){\rule[-0.200pt]{4.818pt}{0.400pt}}
\put(282.0,199.0){\rule[-0.200pt]{0.400pt}{5.059pt}}
\put(272.0,199.0){\rule[-0.200pt]{4.818pt}{0.400pt}}
\put(272.0,220.0){\rule[-0.200pt]{4.818pt}{0.400pt}}
\put(374.0,199.0){\rule[-0.200pt]{0.400pt}{6.263pt}}
\put(364.0,199.0){\rule[-0.200pt]{4.818pt}{0.400pt}}
\put(364.0,225.0){\rule[-0.200pt]{4.818pt}{0.400pt}}
\put(176,187){\usebox{\plotpoint}}
\put(176,187.17){\rule{0.700pt}{0.400pt}}
\multiput(176.00,186.17)(1.547,2.000){2}{\rule{0.350pt}{0.400pt}}
\put(179,188.67){\rule{0.723pt}{0.400pt}}
\multiput(179.00,188.17)(1.500,1.000){2}{\rule{0.361pt}{0.400pt}}
\put(182,189.67){\rule{0.723pt}{0.400pt}}
\multiput(182.00,189.17)(1.500,1.000){2}{\rule{0.361pt}{0.400pt}}
\put(185,190.67){\rule{0.723pt}{0.400pt}}
\multiput(185.00,190.17)(1.500,1.000){2}{\rule{0.361pt}{0.400pt}}
\put(188,191.67){\rule{0.723pt}{0.400pt}}
\multiput(188.00,191.17)(1.500,1.000){2}{\rule{0.361pt}{0.400pt}}
\put(191,192.67){\rule{0.723pt}{0.400pt}}
\multiput(191.00,192.17)(1.500,1.000){2}{\rule{0.361pt}{0.400pt}}
\put(197,193.67){\rule{0.723pt}{0.400pt}}
\multiput(197.00,193.17)(1.500,1.000){2}{\rule{0.361pt}{0.400pt}}
\put(200,194.67){\rule{0.723pt}{0.400pt}}
\multiput(200.00,194.17)(1.500,1.000){2}{\rule{0.361pt}{0.400pt}}
\put(203,195.67){\rule{0.723pt}{0.400pt}}
\multiput(203.00,195.17)(1.500,1.000){2}{\rule{0.361pt}{0.400pt}}
\put(194.0,194.0){\rule[-0.200pt]{0.723pt}{0.400pt}}
\put(209,196.67){\rule{0.723pt}{0.400pt}}
\multiput(209.00,196.17)(1.500,1.000){2}{\rule{0.361pt}{0.400pt}}
\put(212,197.67){\rule{0.723pt}{0.400pt}}
\multiput(212.00,197.17)(1.500,1.000){2}{\rule{0.361pt}{0.400pt}}
\put(206.0,197.0){\rule[-0.200pt]{0.723pt}{0.400pt}}
\put(218,198.67){\rule{0.723pt}{0.400pt}}
\multiput(218.00,198.17)(1.500,1.000){2}{\rule{0.361pt}{0.400pt}}
\put(215.0,199.0){\rule[-0.200pt]{0.723pt}{0.400pt}}
\put(224,199.67){\rule{0.964pt}{0.400pt}}
\multiput(224.00,199.17)(2.000,1.000){2}{\rule{0.482pt}{0.400pt}}
\put(221.0,200.0){\rule[-0.200pt]{0.723pt}{0.400pt}}
\put(231,200.67){\rule{0.723pt}{0.400pt}}
\multiput(231.00,200.17)(1.500,1.000){2}{\rule{0.361pt}{0.400pt}}
\put(228.0,201.0){\rule[-0.200pt]{0.723pt}{0.400pt}}
\put(237,201.67){\rule{0.723pt}{0.400pt}}
\multiput(237.00,201.17)(1.500,1.000){2}{\rule{0.361pt}{0.400pt}}
\put(234.0,202.0){\rule[-0.200pt]{0.723pt}{0.400pt}}
\put(243,202.67){\rule{0.723pt}{0.400pt}}
\multiput(243.00,202.17)(1.500,1.000){2}{\rule{0.361pt}{0.400pt}}
\put(240.0,203.0){\rule[-0.200pt]{0.723pt}{0.400pt}}
\put(252,203.67){\rule{0.723pt}{0.400pt}}
\multiput(252.00,203.17)(1.500,1.000){2}{\rule{0.361pt}{0.400pt}}
\put(246.0,204.0){\rule[-0.200pt]{1.445pt}{0.400pt}}
\put(258,204.67){\rule{0.723pt}{0.400pt}}
\multiput(258.00,204.17)(1.500,1.000){2}{\rule{0.361pt}{0.400pt}}
\put(255.0,205.0){\rule[-0.200pt]{0.723pt}{0.400pt}}
\put(270,205.67){\rule{0.723pt}{0.400pt}}
\multiput(270.00,205.17)(1.500,1.000){2}{\rule{0.361pt}{0.400pt}}
\put(261.0,206.0){\rule[-0.200pt]{2.168pt}{0.400pt}}
\put(279,206.67){\rule{0.723pt}{0.400pt}}
\multiput(279.00,206.17)(1.500,1.000){2}{\rule{0.361pt}{0.400pt}}
\put(273.0,207.0){\rule[-0.200pt]{1.445pt}{0.400pt}}
\put(291,207.67){\rule{0.723pt}{0.400pt}}
\multiput(291.00,207.17)(1.500,1.000){2}{\rule{0.361pt}{0.400pt}}
\put(282.0,208.0){\rule[-0.200pt]{2.168pt}{0.400pt}}
\put(303,208.67){\rule{0.723pt}{0.400pt}}
\multiput(303.00,208.17)(1.500,1.000){2}{\rule{0.361pt}{0.400pt}}
\put(294.0,209.0){\rule[-0.200pt]{2.168pt}{0.400pt}}
\put(318,209.67){\rule{0.723pt}{0.400pt}}
\multiput(318.00,209.17)(1.500,1.000){2}{\rule{0.361pt}{0.400pt}}
\put(306.0,210.0){\rule[-0.200pt]{2.891pt}{0.400pt}}
\put(337,210.67){\rule{0.723pt}{0.400pt}}
\multiput(337.00,210.17)(1.500,1.000){2}{\rule{0.361pt}{0.400pt}}
\put(321.0,211.0){\rule[-0.200pt]{3.854pt}{0.400pt}}
\put(358,211.67){\rule{0.723pt}{0.400pt}}
\multiput(358.00,211.17)(1.500,1.000){2}{\rule{0.361pt}{0.400pt}}
\put(340.0,212.0){\rule[-0.200pt]{4.336pt}{0.400pt}}
\put(382,212.67){\rule{0.723pt}{0.400pt}}
\multiput(382.00,212.17)(1.500,1.000){2}{\rule{0.361pt}{0.400pt}}
\put(361.0,213.0){\rule[-0.200pt]{5.059pt}{0.400pt}}
\put(409,213.67){\rule{0.723pt}{0.400pt}}
\multiput(409.00,213.17)(1.500,1.000){2}{\rule{0.361pt}{0.400pt}}
\put(385.0,214.0){\rule[-0.200pt]{5.782pt}{0.400pt}}
\put(443,214.67){\rule{0.723pt}{0.400pt}}
\multiput(443.00,214.17)(1.500,1.000){2}{\rule{0.361pt}{0.400pt}}
\put(412.0,215.0){\rule[-0.200pt]{7.468pt}{0.400pt}}
\put(446.0,216.0){\rule[-0.200pt]{7.227pt}{0.400pt}}
\end{picture}
\hspace*{-0.8in}
\setlength{\unitlength}{0.240900pt}
\ifx\plotpoint\undefined\newsavebox{\plotpoint}\fi
\sbox{\plotpoint}{\rule[-0.200pt]{0.400pt}{0.400pt}}%
\begin{picture}(540,450)(0,0)
\font\gnuplot=cmr10 at 10pt
\gnuplot
\sbox{\plotpoint}{\rule[-0.200pt]{0.400pt}{0.400pt}}%
\put(176.0,68.0){\rule[-0.200pt]{4.818pt}{0.400pt}}
\put(176.0,120.0){\rule[-0.200pt]{4.818pt}{0.400pt}}
\put(176.0,173.0){\rule[-0.200pt]{4.818pt}{0.400pt}}
\put(176.0,225.0){\rule[-0.200pt]{4.818pt}{0.400pt}}
\put(176.0,277.0){\rule[-0.200pt]{4.818pt}{0.400pt}}
\put(176.0,330.0){\rule[-0.200pt]{4.818pt}{0.400pt}}
\put(176.0,382.0){\rule[-0.200pt]{4.818pt}{0.400pt}}
\put(176.0,68.0){\rule[-0.200pt]{0.400pt}{4.818pt}}
\put(210,23){\makebox(0,0){0}}
\put(176.0,362.0){\rule[-0.200pt]{0.400pt}{4.818pt}}
\put(251.0,68.0){\rule[-0.200pt]{0.400pt}{4.818pt}}
\put(261,23){\makebox(0,0){0.2}}
\put(251.0,362.0){\rule[-0.200pt]{0.400pt}{4.818pt}}
\put(326.0,68.0){\rule[-0.200pt]{0.400pt}{4.818pt}}
\put(326,23){\makebox(0,0){0.4}}
\put(326.0,362.0){\rule[-0.200pt]{0.400pt}{4.818pt}}
\put(401.0,68.0){\rule[-0.200pt]{0.400pt}{4.818pt}}
\put(401,23){\makebox(0,0){0.6}}
\put(401.0,362.0){\rule[-0.200pt]{0.400pt}{4.818pt}}
\put(476.0,68.0){\rule[-0.200pt]{0.400pt}{4.818pt}}
\put(460,23){\makebox(0,0){0.8}}
\put(476.0,362.0){\rule[-0.200pt]{0.400pt}{4.818pt}}
\put(176.0,68.0){\rule[-0.200pt]{72.270pt}{0.400pt}}
\put(476.0,68.0){\rule[-0.200pt]{0.400pt}{75.643pt}}
\put(176.0,382.0){\rule[-0.200pt]{72.270pt}{0.400pt}}
\put(326,337){\makebox(0,0){{$\kappa = 0.152$}}}
\put(326,-20){\makebox(0,0){{$-q^{2}a^{2}$}}}
\put(176.0,68.0){\rule[-0.200pt]{0.400pt}{75.643pt}}
\put(232,183){\circle{12}}
\put(286,191){\circle{12}}
\put(385,209){\circle{12}}
\put(232.0,173.0){\rule[-0.200pt]{0.400pt}{5.059pt}}
\put(222.0,173.0){\rule[-0.200pt]{4.818pt}{0.400pt}}
\put(222.0,194.0){\rule[-0.200pt]{4.818pt}{0.400pt}}
\put(286.0,178.0){\rule[-0.200pt]{0.400pt}{6.263pt}}
\put(276.0,178.0){\rule[-0.200pt]{4.818pt}{0.400pt}}
\put(276.0,204.0){\rule[-0.200pt]{4.818pt}{0.400pt}}
\put(385.0,192.0){\rule[-0.200pt]{0.400pt}{8.191pt}}
\put(375.0,192.0){\rule[-0.200pt]{4.818pt}{0.400pt}}
\put(375.0,226.0){\rule[-0.200pt]{4.818pt}{0.400pt}}
\put(176,153){\usebox{\plotpoint}}
\put(176,153.17){\rule{0.700pt}{0.400pt}}
\multiput(176.00,152.17)(1.547,2.000){2}{\rule{0.350pt}{0.400pt}}
\multiput(179.00,155.61)(0.462,0.447){3}{\rule{0.500pt}{0.108pt}}
\multiput(179.00,154.17)(1.962,3.000){2}{\rule{0.250pt}{0.400pt}}
\put(182,158.17){\rule{0.700pt}{0.400pt}}
\multiput(182.00,157.17)(1.547,2.000){2}{\rule{0.350pt}{0.400pt}}
\put(185,160.17){\rule{0.700pt}{0.400pt}}
\multiput(185.00,159.17)(1.547,2.000){2}{\rule{0.350pt}{0.400pt}}
\put(188,162.17){\rule{0.700pt}{0.400pt}}
\multiput(188.00,161.17)(1.547,2.000){2}{\rule{0.350pt}{0.400pt}}
\put(191,164.17){\rule{0.700pt}{0.400pt}}
\multiput(191.00,163.17)(1.547,2.000){2}{\rule{0.350pt}{0.400pt}}
\put(194,166.17){\rule{0.700pt}{0.400pt}}
\multiput(194.00,165.17)(1.547,2.000){2}{\rule{0.350pt}{0.400pt}}
\put(197,168.17){\rule{0.700pt}{0.400pt}}
\multiput(197.00,167.17)(1.547,2.000){2}{\rule{0.350pt}{0.400pt}}
\put(200,169.67){\rule{0.723pt}{0.400pt}}
\multiput(200.00,169.17)(1.500,1.000){2}{\rule{0.361pt}{0.400pt}}
\put(203,171.17){\rule{0.700pt}{0.400pt}}
\multiput(203.00,170.17)(1.547,2.000){2}{\rule{0.350pt}{0.400pt}}
\put(206,172.67){\rule{0.723pt}{0.400pt}}
\multiput(206.00,172.17)(1.500,1.000){2}{\rule{0.361pt}{0.400pt}}
\put(209,174.17){\rule{0.700pt}{0.400pt}}
\multiput(209.00,173.17)(1.547,2.000){2}{\rule{0.350pt}{0.400pt}}
\put(212,175.67){\rule{0.723pt}{0.400pt}}
\multiput(212.00,175.17)(1.500,1.000){2}{\rule{0.361pt}{0.400pt}}
\put(215,176.67){\rule{0.723pt}{0.400pt}}
\multiput(215.00,176.17)(1.500,1.000){2}{\rule{0.361pt}{0.400pt}}
\put(218,177.67){\rule{0.723pt}{0.400pt}}
\multiput(218.00,177.17)(1.500,1.000){2}{\rule{0.361pt}{0.400pt}}
\put(221,178.67){\rule{0.723pt}{0.400pt}}
\multiput(221.00,178.17)(1.500,1.000){2}{\rule{0.361pt}{0.400pt}}
\put(224,179.67){\rule{0.964pt}{0.400pt}}
\multiput(224.00,179.17)(2.000,1.000){2}{\rule{0.482pt}{0.400pt}}
\put(228,180.67){\rule{0.723pt}{0.400pt}}
\multiput(228.00,180.17)(1.500,1.000){2}{\rule{0.361pt}{0.400pt}}
\put(231,181.67){\rule{0.723pt}{0.400pt}}
\multiput(231.00,181.17)(1.500,1.000){2}{\rule{0.361pt}{0.400pt}}
\put(234,182.67){\rule{0.723pt}{0.400pt}}
\multiput(234.00,182.17)(1.500,1.000){2}{\rule{0.361pt}{0.400pt}}
\put(237,183.67){\rule{0.723pt}{0.400pt}}
\multiput(237.00,183.17)(1.500,1.000){2}{\rule{0.361pt}{0.400pt}}
\put(240,184.67){\rule{0.723pt}{0.400pt}}
\multiput(240.00,184.17)(1.500,1.000){2}{\rule{0.361pt}{0.400pt}}
\put(243,185.67){\rule{0.723pt}{0.400pt}}
\multiput(243.00,185.17)(1.500,1.000){2}{\rule{0.361pt}{0.400pt}}
\put(249,186.67){\rule{0.723pt}{0.400pt}}
\multiput(249.00,186.17)(1.500,1.000){2}{\rule{0.361pt}{0.400pt}}
\put(252,187.67){\rule{0.723pt}{0.400pt}}
\multiput(252.00,187.17)(1.500,1.000){2}{\rule{0.361pt}{0.400pt}}
\put(255,188.67){\rule{0.723pt}{0.400pt}}
\multiput(255.00,188.17)(1.500,1.000){2}{\rule{0.361pt}{0.400pt}}
\put(246.0,187.0){\rule[-0.200pt]{0.723pt}{0.400pt}}
\put(261,189.67){\rule{0.723pt}{0.400pt}}
\multiput(261.00,189.17)(1.500,1.000){2}{\rule{0.361pt}{0.400pt}}
\put(258.0,190.0){\rule[-0.200pt]{0.723pt}{0.400pt}}
\put(267,190.67){\rule{0.723pt}{0.400pt}}
\multiput(267.00,190.17)(1.500,1.000){2}{\rule{0.361pt}{0.400pt}}
\put(270,191.67){\rule{0.723pt}{0.400pt}}
\multiput(270.00,191.17)(1.500,1.000){2}{\rule{0.361pt}{0.400pt}}
\put(264.0,191.0){\rule[-0.200pt]{0.723pt}{0.400pt}}
\put(276,192.67){\rule{0.723pt}{0.400pt}}
\multiput(276.00,192.17)(1.500,1.000){2}{\rule{0.361pt}{0.400pt}}
\put(273.0,193.0){\rule[-0.200pt]{0.723pt}{0.400pt}}
\put(282,193.67){\rule{0.723pt}{0.400pt}}
\multiput(282.00,193.17)(1.500,1.000){2}{\rule{0.361pt}{0.400pt}}
\put(279.0,194.0){\rule[-0.200pt]{0.723pt}{0.400pt}}
\put(288,194.67){\rule{0.723pt}{0.400pt}}
\multiput(288.00,194.17)(1.500,1.000){2}{\rule{0.361pt}{0.400pt}}
\put(285.0,195.0){\rule[-0.200pt]{0.723pt}{0.400pt}}
\put(294,195.67){\rule{0.723pt}{0.400pt}}
\multiput(294.00,195.17)(1.500,1.000){2}{\rule{0.361pt}{0.400pt}}
\put(291.0,196.0){\rule[-0.200pt]{0.723pt}{0.400pt}}
\put(303,196.67){\rule{0.723pt}{0.400pt}}
\multiput(303.00,196.17)(1.500,1.000){2}{\rule{0.361pt}{0.400pt}}
\put(297.0,197.0){\rule[-0.200pt]{1.445pt}{0.400pt}}
\put(309,197.67){\rule{0.723pt}{0.400pt}}
\multiput(309.00,197.17)(1.500,1.000){2}{\rule{0.361pt}{0.400pt}}
\put(306.0,198.0){\rule[-0.200pt]{0.723pt}{0.400pt}}
\put(318,198.67){\rule{0.723pt}{0.400pt}}
\multiput(318.00,198.17)(1.500,1.000){2}{\rule{0.361pt}{0.400pt}}
\put(312.0,199.0){\rule[-0.200pt]{1.445pt}{0.400pt}}
\put(328,199.67){\rule{0.723pt}{0.400pt}}
\multiput(328.00,199.17)(1.500,1.000){2}{\rule{0.361pt}{0.400pt}}
\put(321.0,200.0){\rule[-0.200pt]{1.686pt}{0.400pt}}
\put(340,200.67){\rule{0.723pt}{0.400pt}}
\multiput(340.00,200.17)(1.500,1.000){2}{\rule{0.361pt}{0.400pt}}
\put(331.0,201.0){\rule[-0.200pt]{2.168pt}{0.400pt}}
\put(349,201.67){\rule{0.723pt}{0.400pt}}
\multiput(349.00,201.17)(1.500,1.000){2}{\rule{0.361pt}{0.400pt}}
\put(343.0,202.0){\rule[-0.200pt]{1.445pt}{0.400pt}}
\put(361,202.67){\rule{0.723pt}{0.400pt}}
\multiput(361.00,202.17)(1.500,1.000){2}{\rule{0.361pt}{0.400pt}}
\put(352.0,203.0){\rule[-0.200pt]{2.168pt}{0.400pt}}
\put(373,203.67){\rule{0.723pt}{0.400pt}}
\multiput(373.00,203.17)(1.500,1.000){2}{\rule{0.361pt}{0.400pt}}
\put(364.0,204.0){\rule[-0.200pt]{2.168pt}{0.400pt}}
\put(388,204.67){\rule{0.723pt}{0.400pt}}
\multiput(388.00,204.17)(1.500,1.000){2}{\rule{0.361pt}{0.400pt}}
\put(376.0,205.0){\rule[-0.200pt]{2.891pt}{0.400pt}}
\put(403,205.67){\rule{0.723pt}{0.400pt}}
\multiput(403.00,205.17)(1.500,1.000){2}{\rule{0.361pt}{0.400pt}}
\put(391.0,206.0){\rule[-0.200pt]{2.891pt}{0.400pt}}
\put(421,206.67){\rule{0.723pt}{0.400pt}}
\multiput(421.00,206.17)(1.500,1.000){2}{\rule{0.361pt}{0.400pt}}
\put(406.0,207.0){\rule[-0.200pt]{3.613pt}{0.400pt}}
\put(440,207.67){\rule{0.723pt}{0.400pt}}
\multiput(440.00,207.17)(1.500,1.000){2}{\rule{0.361pt}{0.400pt}}
\put(424.0,208.0){\rule[-0.200pt]{3.854pt}{0.400pt}}
\put(464,208.67){\rule{0.723pt}{0.400pt}}
\multiput(464.00,208.17)(1.500,1.000){2}{\rule{0.361pt}{0.400pt}}
\put(443.0,209.0){\rule[-0.200pt]{5.059pt}{0.400pt}}
\put(467.0,210.0){\rule[-0.200pt]{2.168pt}{0.400pt}}
\end{picture}
\hspace*{-0.8in}
\setlength{\unitlength}{0.240900pt}
\ifx\plotpoint\undefined\newsavebox{\plotpoint}\fi
\begin{picture}(540,450)(0,0)
\font\gnuplot=cmr10 at 10pt
\gnuplot
\sbox{\plotpoint}{\rule[-0.200pt]{0.400pt}{0.400pt}}%
\put(176.0,68.0){\rule[-0.200pt]{4.818pt}{0.400pt}}
\put(176.0,120.0){\rule[-0.200pt]{4.818pt}{0.400pt}}
\put(176.0,173.0){\rule[-0.200pt]{4.818pt}{0.400pt}}
\put(176.0,225.0){\rule[-0.200pt]{4.818pt}{0.400pt}}
\put(456.0,225.0){\rule[-0.200pt]{4.818pt}{0.400pt}}
\put(176.0,277.0){\rule[-0.200pt]{4.818pt}{0.400pt}}
\put(176.0,330.0){\rule[-0.200pt]{4.818pt}{0.400pt}}
\put(176.0,382.0){\rule[-0.200pt]{4.818pt}{0.400pt}}
\put(176.0,68.0){\rule[-0.200pt]{0.400pt}{4.818pt}}
\put(210,23){\makebox(0,0){0}}
\put(176.0,362.0){\rule[-0.200pt]{0.400pt}{4.818pt}}
\put(251.0,68.0){\rule[-0.200pt]{0.400pt}{4.818pt}}
\put(261,23){\makebox(0,0){0.2}}
\put(251.0,362.0){\rule[-0.200pt]{0.400pt}{4.818pt}}
\put(326.0,68.0){\rule[-0.200pt]{0.400pt}{4.818pt}}
\put(326,23){\makebox(0,0){0.4}}
\put(326.0,362.0){\rule[-0.200pt]{0.400pt}{4.818pt}}
\put(401.0,68.0){\rule[-0.200pt]{0.400pt}{4.818pt}}
\put(391,23){\makebox(0,0){0.6}}
\put(401.0,362.0){\rule[-0.200pt]{0.400pt}{4.818pt}}
\put(476.0,68.0){\rule[-0.200pt]{0.400pt}{4.818pt}}
\put(456,23){\makebox(0,0){0.8}}
\put(476.0,362.0){\rule[-0.200pt]{0.400pt}{4.818pt}}
\put(176.0,68.0){\rule[-0.200pt]{72.270pt}{0.400pt}}
\put(476.0,68.0){\rule[-0.200pt]{0.400pt}{75.643pt}}
\put(176.0,382.0){\rule[-0.200pt]{72.270pt}{0.400pt}}
\put(326,337){\makebox(0,0){{$\kappa = 0.154$}}}
\put(176.0,68.0){\rule[-0.200pt]{0.400pt}{75.643pt}}
\put(230,181){\circle{12}}
\put(279,194){\circle{12}}
\put(363,207){\circle{12}}
\put(230.0,167.0){\rule[-0.200pt]{0.400pt}{6.504pt}}
\put(220.0,167.0){\rule[-0.200pt]{4.818pt}{0.400pt}}
\put(220.0,194.0){\rule[-0.200pt]{4.818pt}{0.400pt}}
\put(279.0,178.0){\rule[-0.200pt]{0.400pt}{7.468pt}}
\put(269.0,178.0){\rule[-0.200pt]{4.818pt}{0.400pt}}
\put(269.0,209.0){\rule[-0.200pt]{4.818pt}{0.400pt}}
\put(363.0,187.0){\rule[-0.200pt]{0.400pt}{9.395pt}}
\put(353.0,187.0){\rule[-0.200pt]{4.818pt}{0.400pt}}
\put(353.0,226.0){\rule[-0.200pt]{4.818pt}{0.400pt}}
\put(176,132){\usebox{\plotpoint}}
\multiput(176.61,132.00)(0.447,1.132){3}{\rule{0.108pt}{0.900pt}}
\multiput(175.17,132.00)(3.000,4.132){2}{\rule{0.400pt}{0.450pt}}
\multiput(179.61,138.00)(0.447,0.685){3}{\rule{0.108pt}{0.633pt}}
\multiput(178.17,138.00)(3.000,2.685){2}{\rule{0.400pt}{0.317pt}}
\multiput(182.61,142.00)(0.447,0.909){3}{\rule{0.108pt}{0.767pt}}
\multiput(181.17,142.00)(3.000,3.409){2}{\rule{0.400pt}{0.383pt}}
\multiput(185.00,147.61)(0.462,0.447){3}{\rule{0.500pt}{0.108pt}}
\multiput(185.00,146.17)(1.962,3.000){2}{\rule{0.250pt}{0.400pt}}
\multiput(188.61,150.00)(0.447,0.685){3}{\rule{0.108pt}{0.633pt}}
\multiput(187.17,150.00)(3.000,2.685){2}{\rule{0.400pt}{0.317pt}}
\multiput(191.00,154.61)(0.462,0.447){3}{\rule{0.500pt}{0.108pt}}
\multiput(191.00,153.17)(1.962,3.000){2}{\rule{0.250pt}{0.400pt}}
\multiput(194.00,157.61)(0.462,0.447){3}{\rule{0.500pt}{0.108pt}}
\multiput(194.00,156.17)(1.962,3.000){2}{\rule{0.250pt}{0.400pt}}
\put(197,160.17){\rule{0.700pt}{0.400pt}}
\multiput(197.00,159.17)(1.547,2.000){2}{\rule{0.350pt}{0.400pt}}
\multiput(200.00,162.61)(0.462,0.447){3}{\rule{0.500pt}{0.108pt}}
\multiput(200.00,161.17)(1.962,3.000){2}{\rule{0.250pt}{0.400pt}}
\put(203,165.17){\rule{0.700pt}{0.400pt}}
\multiput(203.00,164.17)(1.547,2.000){2}{\rule{0.350pt}{0.400pt}}
\put(206,167.17){\rule{0.700pt}{0.400pt}}
\multiput(206.00,166.17)(1.547,2.000){2}{\rule{0.350pt}{0.400pt}}
\put(209,169.17){\rule{0.700pt}{0.400pt}}
\multiput(209.00,168.17)(1.547,2.000){2}{\rule{0.350pt}{0.400pt}}
\put(212,171.17){\rule{0.700pt}{0.400pt}}
\multiput(212.00,170.17)(1.547,2.000){2}{\rule{0.350pt}{0.400pt}}
\put(215,173.17){\rule{0.700pt}{0.400pt}}
\multiput(215.00,172.17)(1.547,2.000){2}{\rule{0.350pt}{0.400pt}}
\put(218,174.67){\rule{0.723pt}{0.400pt}}
\multiput(218.00,174.17)(1.500,1.000){2}{\rule{0.361pt}{0.400pt}}
\put(221,176.17){\rule{0.700pt}{0.400pt}}
\multiput(221.00,175.17)(1.547,2.000){2}{\rule{0.350pt}{0.400pt}}
\put(224,177.67){\rule{0.964pt}{0.400pt}}
\multiput(224.00,177.17)(2.000,1.000){2}{\rule{0.482pt}{0.400pt}}
\put(228,178.67){\rule{0.723pt}{0.400pt}}
\multiput(228.00,178.17)(1.500,1.000){2}{\rule{0.361pt}{0.400pt}}
\put(231,180.17){\rule{0.700pt}{0.400pt}}
\multiput(231.00,179.17)(1.547,2.000){2}{\rule{0.350pt}{0.400pt}}
\put(234,181.67){\rule{0.723pt}{0.400pt}}
\multiput(234.00,181.17)(1.500,1.000){2}{\rule{0.361pt}{0.400pt}}
\put(237,182.67){\rule{0.723pt}{0.400pt}}
\multiput(237.00,182.17)(1.500,1.000){2}{\rule{0.361pt}{0.400pt}}
\put(240,183.67){\rule{0.723pt}{0.400pt}}
\multiput(240.00,183.17)(1.500,1.000){2}{\rule{0.361pt}{0.400pt}}
\put(243,184.67){\rule{0.723pt}{0.400pt}}
\multiput(243.00,184.17)(1.500,1.000){2}{\rule{0.361pt}{0.400pt}}
\put(246,185.67){\rule{0.723pt}{0.400pt}}
\multiput(246.00,185.17)(1.500,1.000){2}{\rule{0.361pt}{0.400pt}}
\put(249,186.67){\rule{0.723pt}{0.400pt}}
\multiput(249.00,186.17)(1.500,1.000){2}{\rule{0.361pt}{0.400pt}}
\put(252,187.67){\rule{0.723pt}{0.400pt}}
\multiput(252.00,187.17)(1.500,1.000){2}{\rule{0.361pt}{0.400pt}}
\put(255,188.67){\rule{0.723pt}{0.400pt}}
\multiput(255.00,188.17)(1.500,1.000){2}{\rule{0.361pt}{0.400pt}}
\put(261,189.67){\rule{0.723pt}{0.400pt}}
\multiput(261.00,189.17)(1.500,1.000){2}{\rule{0.361pt}{0.400pt}}
\put(264,190.67){\rule{0.723pt}{0.400pt}}
\multiput(264.00,190.17)(1.500,1.000){2}{\rule{0.361pt}{0.400pt}}
\put(267,191.67){\rule{0.723pt}{0.400pt}}
\multiput(267.00,191.17)(1.500,1.000){2}{\rule{0.361pt}{0.400pt}}
\put(258.0,190.0){\rule[-0.200pt]{0.723pt}{0.400pt}}
\put(273,192.67){\rule{0.723pt}{0.400pt}}
\multiput(273.00,192.17)(1.500,1.000){2}{\rule{0.361pt}{0.400pt}}
\put(276,193.67){\rule{0.723pt}{0.400pt}}
\multiput(276.00,193.17)(1.500,1.000){2}{\rule{0.361pt}{0.400pt}}
\put(270.0,193.0){\rule[-0.200pt]{0.723pt}{0.400pt}}
\put(282,194.67){\rule{0.723pt}{0.400pt}}
\multiput(282.00,194.17)(1.500,1.000){2}{\rule{0.361pt}{0.400pt}}
\put(279.0,195.0){\rule[-0.200pt]{0.723pt}{0.400pt}}
\put(288,195.67){\rule{0.723pt}{0.400pt}}
\multiput(288.00,195.17)(1.500,1.000){2}{\rule{0.361pt}{0.400pt}}
\put(285.0,196.0){\rule[-0.200pt]{0.723pt}{0.400pt}}
\put(294,196.67){\rule{0.723pt}{0.400pt}}
\multiput(294.00,196.17)(1.500,1.000){2}{\rule{0.361pt}{0.400pt}}
\put(291.0,197.0){\rule[-0.200pt]{0.723pt}{0.400pt}}
\put(300,197.67){\rule{0.723pt}{0.400pt}}
\multiput(300.00,197.17)(1.500,1.000){2}{\rule{0.361pt}{0.400pt}}
\put(297.0,198.0){\rule[-0.200pt]{0.723pt}{0.400pt}}
\put(306,198.67){\rule{0.723pt}{0.400pt}}
\multiput(306.00,198.17)(1.500,1.000){2}{\rule{0.361pt}{0.400pt}}
\put(303.0,199.0){\rule[-0.200pt]{0.723pt}{0.400pt}}
\put(315,199.67){\rule{0.723pt}{0.400pt}}
\multiput(315.00,199.17)(1.500,1.000){2}{\rule{0.361pt}{0.400pt}}
\put(309.0,200.0){\rule[-0.200pt]{1.445pt}{0.400pt}}
\put(321,200.67){\rule{0.723pt}{0.400pt}}
\multiput(321.00,200.17)(1.500,1.000){2}{\rule{0.361pt}{0.400pt}}
\put(318.0,201.0){\rule[-0.200pt]{0.723pt}{0.400pt}}
\put(331,201.67){\rule{0.723pt}{0.400pt}}
\multiput(331.00,201.17)(1.500,1.000){2}{\rule{0.361pt}{0.400pt}}
\put(324.0,202.0){\rule[-0.200pt]{1.686pt}{0.400pt}}
\put(343,202.67){\rule{0.723pt}{0.400pt}}
\multiput(343.00,202.17)(1.500,1.000){2}{\rule{0.361pt}{0.400pt}}
\put(334.0,203.0){\rule[-0.200pt]{2.168pt}{0.400pt}}
\put(352,203.67){\rule{0.723pt}{0.400pt}}
\multiput(352.00,203.17)(1.500,1.000){2}{\rule{0.361pt}{0.400pt}}
\put(346.0,204.0){\rule[-0.200pt]{1.445pt}{0.400pt}}
\put(364,204.67){\rule{0.723pt}{0.400pt}}
\multiput(364.00,204.17)(1.500,1.000){2}{\rule{0.361pt}{0.400pt}}
\put(355.0,205.0){\rule[-0.200pt]{2.168pt}{0.400pt}}
\put(376,205.67){\rule{0.723pt}{0.400pt}}
\multiput(376.00,205.17)(1.500,1.000){2}{\rule{0.361pt}{0.400pt}}
\put(367.0,206.0){\rule[-0.200pt]{2.168pt}{0.400pt}}
\put(391,206.67){\rule{0.723pt}{0.400pt}}
\multiput(391.00,206.17)(1.500,1.000){2}{\rule{0.361pt}{0.400pt}}
\put(379.0,207.0){\rule[-0.200pt]{2.891pt}{0.400pt}}
\put(406,207.67){\rule{0.723pt}{0.400pt}}
\multiput(406.00,207.17)(1.500,1.000){2}{\rule{0.361pt}{0.400pt}}
\put(394.0,208.0){\rule[-0.200pt]{2.891pt}{0.400pt}}
\put(424,208.67){\rule{0.964pt}{0.400pt}}
\multiput(424.00,208.17)(2.000,1.000){2}{\rule{0.482pt}{0.400pt}}
\put(409.0,209.0){\rule[-0.200pt]{3.613pt}{0.400pt}}
\put(446,209.67){\rule{0.723pt}{0.400pt}}
\multiput(446.00,209.17)(1.500,1.000){2}{\rule{0.361pt}{0.400pt}}
\put(428.0,210.0){\rule[-0.200pt]{4.336pt}{0.400pt}}
\put(470,210.67){\rule{0.723pt}{0.400pt}}
\multiput(470.00,210.17)(1.500,1.000){2}{\rule{0.361pt}{0.400pt}}
\put(449.0,211.0){\rule[-0.200pt]{5.059pt}{0.400pt}}
\put(473.0,212.0){\rule[-0.200pt]{0.723pt}{0.400pt}}
\end{picture}
\]
\vspace*{-0.1in}
\caption{Monopole fitting for $T_{DI}(q^2)$. $T_{DI}(0)$ values 
are obtained by extrapolating to $q^{2}\rightarrow 0$.}
\end{figure}
\vspace{-0.08in}
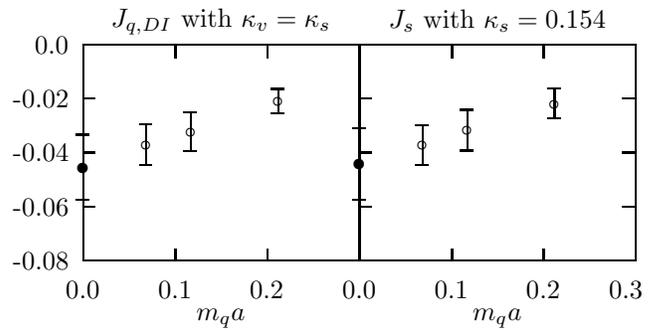
\begin{figure}[h]
\[
\hspace*{-0.14in}
\setlength{\unitlength}{0.240900pt}
\ifx\plotpoint\undefined\newsavebox{\plotpoint}\fi
\begin{picture}(674,521)(0,0)
\font\gnuplot=cmr10 at 10pt
\gnuplot
\sbox{\plotpoint}{\rule[-0.200pt]{0.400pt}{0.400pt}}%
\put(176.0,113.0){\rule[-0.200pt]{4.818pt}{0.400pt}}
\put(154,113){\makebox(0,0)[r]{-0.08}}
\put(176.0,198.0){\rule[-0.200pt]{4.818pt}{0.400pt}}
\put(154,198){\makebox(0,0)[r]{-0.06}}
\put(176.0,283.0){\rule[-0.200pt]{4.818pt}{0.400pt}}
\put(154,283){\makebox(0,0)[r]{-0.04}}
\put(176.0,368.0){\rule[-0.200pt]{4.818pt}{0.400pt}}
\put(154,368){\makebox(0,0)[r]{-0.02}}
\put(176.0,453.0){\rule[-0.200pt]{4.818pt}{0.400pt}}
\put(154,453){\makebox(0,0)[r]{0.0}}
\put(176.0,113.0){\rule[-0.200pt]{0.400pt}{4.818pt}}
\put(176,68){\makebox(0,0){0.0}}
\put(176.0,433.0){\rule[-0.200pt]{0.400pt}{4.818pt}}
\put(321.0,113.0){\rule[-0.200pt]{0.400pt}{4.818pt}}
\put(321,68){\makebox(0,0){0.1}}
\put(321.0,433.0){\rule[-0.200pt]{0.400pt}{4.818pt}}
\put(465.0,113.0){\rule[-0.200pt]{0.400pt}{4.818pt}}
\put(465,68){\makebox(0,0){0.2}}
\put(465.0,433.0){\rule[-0.200pt]{0.400pt}{4.818pt}}
\put(610.0,113.0){\rule[-0.200pt]{0.400pt}{4.818pt}}
\put(610.0,433.0){\rule[-0.200pt]{0.400pt}{4.818pt}}
\put(176.0,113.0){\rule[-0.200pt]{104.551pt}{0.400pt}}
\put(610.0,113.0){\rule[-0.200pt]{0.400pt}{81.906pt}}
\put(176.0,453.0){\rule[-0.200pt]{104.551pt}{0.400pt}}
\put(393,23){\makebox(0,0){{$m_{q}a$}}}
\put(393,-25){\makebox(0,0){{$(a)$}}}
\put(393,490){\makebox(0,0){{$J_{q,DI}$ with $\kappa_v = \kappa_s$}}}
\put(176.0,113.0){\rule[-0.200pt]{0.400pt}{81.906pt}}
\put(176,260){\circle*{16}}
\put(275,295){\circle{12}}
\put(345,316){\circle{12}}
\put(482,364){\circle{12}}
\put(176.0,209.0){\rule[-0.200pt]{0.400pt}{24.572pt}}
\put(166.0,209.0){\rule[-0.200pt]{4.818pt}{0.400pt}}
\put(166.0,311.0){\rule[-0.200pt]{4.818pt}{0.400pt}}
\put(275.0,263.0){\rule[-0.200pt]{0.400pt}{15.418pt}}
\put(265.0,263.0){\rule[-0.200pt]{4.818pt}{0.400pt}}
\put(265.0,327.0){\rule[-0.200pt]{4.818pt}{0.400pt}}
\put(345.0,285.0){\rule[-0.200pt]{0.400pt}{14.695pt}}
\put(335.0,285.0){\rule[-0.200pt]{4.818pt}{0.400pt}}
\put(335.0,346.0){\rule[-0.200pt]{4.818pt}{0.400pt}}
\put(482.0,345.0){\rule[-0.200pt]{0.400pt}{9.154pt}}
\put(472.0,345.0){\rule[-0.200pt]{4.818pt}{0.400pt}}
\put(472.0,383.0){\rule[-0.200pt]{4.818pt}{0.400pt}}
\end{picture}
\hspace*{-0.8in}
\setlength{\unitlength}{0.240900pt}
\ifx\plotpoint\undefined\newsavebox{\plotpoint}\fi
\sbox{\plotpoint}{\rule[-0.200pt]{0.400pt}{0.400pt}}%
\begin{picture}(674,521)(0,0)
\font\gnuplot=cmr10 at 10pt
\gnuplot
\sbox{\plotpoint}{\rule[-0.200pt]{0.400pt}{0.400pt}}%
\put(176.0,113.0){\rule[-0.200pt]{4.818pt}{0.400pt}}
\put(590.0,113.0){\rule[-0.200pt]{4.818pt}{0.400pt}}
\put(176.0,198.0){\rule[-0.200pt]{4.818pt}{0.400pt}}
\put(590.0,198.0){\rule[-0.200pt]{4.818pt}{0.400pt}}
\put(176.0,283.0){\rule[-0.200pt]{4.818pt}{0.400pt}}
\put(590.0,283.0){\rule[-0.200pt]{4.818pt}{0.400pt}}
\put(176.0,368.0){\rule[-0.200pt]{4.818pt}{0.400pt}}
\put(590.0,368.0){\rule[-0.200pt]{4.818pt}{0.400pt}}
\put(176.0,453.0){\rule[-0.200pt]{4.818pt}{0.400pt}}
\put(590.0,453.0){\rule[-0.200pt]{4.818pt}{0.400pt}}
\put(176.0,113.0){\rule[-0.200pt]{0.400pt}{4.818pt}}
\put(176,68){\makebox(0,0){0.0}}
\put(176.0,433.0){\rule[-0.200pt]{0.400pt}{4.818pt}}
\put(321.0,113.0){\rule[-0.200pt]{0.400pt}{4.818pt}}
\put(321,68){\makebox(0,0){0.1}}
\put(321.0,433.0){\rule[-0.200pt]{0.400pt}{4.818pt}}
\put(465.0,113.0){\rule[-0.200pt]{0.400pt}{4.818pt}}
\put(465,68){\makebox(0,0){0.2}}
\put(465.0,433.0){\rule[-0.200pt]{0.400pt}{4.818pt}}
\put(610.0,113.0){\rule[-0.200pt]{0.400pt}{4.818pt}}
\put(595,68){\makebox(0,0){0.3}}
\put(610.0,433.0){\rule[-0.200pt]{0.400pt}{4.818pt}}
\put(176.0,113.0){\rule[-0.200pt]{104.551pt}{0.400pt}}
\put(610.0,113.0){\rule[-0.200pt]{0.400pt}{81.906pt}}
\put(176.0,453.0){\rule[-0.200pt]{104.551pt}{0.400pt}}
\put(393,23){\makebox(0,0){{$m_{q}a$}}}
\put(393,-25){\makebox(0,0){{$(b)$}}}
\put(393,490){\makebox(0,0){{$J_s$ with $\kappa_s = 0.154$}}}
\put(176.0,113.0){\rule[-0.200pt]{0.400pt}{81.906pt}}
\put(176,265){\circle*{16}}
\put(275,295){\circle{12}}
\put(345,318){\circle{12}}
\put(482,360){\circle{12}}
\put(176.0,209.0){\rule[-0.200pt]{0.400pt}{26.981pt}}
\put(166.0,209.0){\rule[-0.200pt]{4.818pt}{0.400pt}}
\put(166.0,321.0){\rule[-0.200pt]{4.818pt}{0.400pt}}
\put(275.0,263.0){\rule[-0.200pt]{0.400pt}{15.177pt}}
\put(265.0,263.0){\rule[-0.200pt]{4.818pt}{0.400pt}}
\put(265.0,326.0){\rule[-0.200pt]{4.818pt}{0.400pt}}
\put(345.0,286.0){\rule[-0.200pt]{0.400pt}{15.418pt}}
\put(335.0,286.0){\rule[-0.200pt]{4.818pt}{0.400pt}}
\put(335.0,350.0){\rule[-0.200pt]{4.818pt}{0.400pt}}
\put(482.0,337.0){\rule[-0.200pt]{0.400pt}{11.322pt}}
\put(472.0,337.0){\rule[-0.200pt]{4.818pt}{0.400pt}}
\put(472.0,384.0){\rule[-0.200pt]{4.818pt}{0.400pt}}
\end{picture}
\]
\caption{$(a)$ The lattice $J_{q, DI}$ as a function of the quark mass 
$m_{q}a$. The quark masses in the valence and the sea are kept the same. 
$(b)$ $J_s$ vs the valence quark mass with the sea quark mass fixed at
$\kappa_s = 0.154$. The chiral limit value ($\kappa_v = \kappa_c$) is \
indicated by $\bullet$.} 
\end{figure}
This
is reminiscent of the sea-quark contribution in the flavor-singlet
$g_A^0$ calculation~\cite{dll95} where the sea-flavor independence was
first observed. 

The breakdown of the quark angular momentum $J_q$ into the quark spin
${1 \over 2}\Sigma$ and the quark orbital angular momentum is given in
Table 1.
From the CI calculation we obtain the valence and cloud quark contributions to
the quark angular momentum $J_{q, CI} = 0.44 \pm 0.07$ which is 
$\sim 90\%$ of the total proton spin and it almost saturates the spin 
sum rule in Eq. (\ref{ssr}) by itself. 
A previous calculation of the flavor-singlet 
axial current on the same set of lattices shows that 
${1\over 2} \Sigma_{CI} = 0.31 \pm 0.04$~\cite{dll95}. Since 
$J_{q, CI} = {1\over 2} \Sigma_{CI} + L_{q, CI}$, we obtain the CI part of
the quark orbital angular momentum $L_{q, CI} = 0.13 \pm 0.07$. Thus,
for valence and cloud quarks, about 70\% of $J_{q, CI}$ comes from the 
quark spin and the 30\% is due to the orbital angular momentum. 
From the DI calculation, we find that the total quark angular momentum 
$J_{q, DI}$, like the quark spin ${1\over 2} \Sigma_{DI}$, is also flavor 
symmetric within errors. In fact, $J_{u, DI}, J_{d, DI}$, and
$ J_s$ are all equal to $ -0.047 \pm 0.012$. Together, the total DI is
$J_{q, DI} = -0.14 \pm 0.04$. Subtracting the DI of the quark spin
${1\over 2} \Sigma_{DI} = -0.18 \pm 0.03$ from $J_{q, DI}$, we obtain 
the orbital angular momentum contribution from the sea quarks to be
$L_{q, DI} = 0.041 \pm 0.035$.
It is interesting to note that it is consistent with zero with a
central value which is a factor
of 4.5 smaller than the spin content of the sea quarks. 
Adding CI and DI contributions together, we obtain 
$J_{q} = 0.30 \pm 0.07$ and, thus, we predict the gluon angular
momentum
$J_{g} = 0.5 - 0.30 \pm 0.07 = 0.20 \pm 0.07$ from the spin sum rule
(Eq. (\ref{ssr})). 

To conclude, the total angular momentum of the quarks is calculated to be 
$J_{q} = 0.30 \pm 0.07$, {\it i. e.} 
$\sim 60\%$ of the proton spin is attributable to the quarks.
Since the quark spin content is calculated previously to be ${1 \over 2} 
\Sigma = 0.13 \pm 0.06$~\cite{dll95}, we obtain the quark orbital
angular momentum $L_q = 0. 17 \pm 0.06$. Therefore, about 25\% of the
proton spin originates from the quark spin and about 35\% comes from the 
quark orbital angular momentum. 
The gluon angular momentum contribution is predicted from the spin sum rule
to be  $J_{g} = 0.20 \pm 0.07$, \mbox{{\it i. e.}} $\sim$ 40\% of 
the proton spin is due to the glue.
Since the orbital angular momentum of the sea quarks turns out to be 
quite small, the sea flavor independence of $J_{q, DI}$ reconfirms the
sea flavor independence of the quark spin, namely  $\Delta u (DI) = 
\Delta d (DI) \simeq \Delta s$ observed in the previous lattice calculation
~\cite{dll95}.

\begin{table}
\begin{center}
\caption{The Breakup of Quark Angular Momentum}
\begin{tabular}{cccc}
&$J_{q}$&$\frac{1}{2}\Sigma$&$L_{q}$\\
\hline
$u + d (CI)$&$0.44(7)$&$0.31(4)$& 0.13(7)\\
$u/d (DI) $&$-0.047(12)$&$-0.062(6)$& 0.015(12)\\
$s$&$-0.047(12)$&$-0.058(6)$& 0.011(12)\\
$u + d +s (DI)$& $ - 0.14(4)$& $ - 0.18(3)$& 0.041(36) \\
\hline
Total&$0.30(7)$&$0.13(6)$&$0.17(6)$\\
\end{tabular}
\end{center}
\end{table}

\noindent
In addition, the fact that $J_{q, CI}$ almost saturates the
spin sum rule and the sea quark orbital angular momentum is small,
the gluon angular momentum and the sea quark spin largely cancel each other.
Both the flavor independence in the sea and the cancellation between the gluon 
angular momentum and the sea quark spin suggest that it is related to the
anomaly and anomalous chiral Ward identity.

It is important for the future experiments to measure the quark
orbital angular momentum and the gluon angular momentum in order to
conclude the study of the spin structure and content of the proton. 
The present work is 
based on the quenched approximation and is subject to the large volume,
finite lattice spacing and chiral extrapolation corrections. We shall 
address these issues in a future study with the overlap fermion~\cite{neu98}
which has nice chiral and scaling properties.

This work is partially supported by DOE Grant No. DE-FGO5-84ER40154, 
DE-FG02-88ER40448, NSF Grant No. 9722073, and BMBF. We thank X. Ji for 
fruitful discussions.

\end{document}